\documentclass[fleqn,usenatbib]{mnras}

\usepackage{newtxtext,newtxmath}

\usepackage[T1]{fontenc}

\DeclareRobustCommand{\VAN}[3]{#2}
\let\VANthebibliography\thebibliography
\def\thebibliography{\DeclareRobustCommand{\VAN}[3]{##3}\VANthebibliography}

\usepackage{makecell}
\usepackage{siunitx}
\usepackage{bm}

\usepackage{graphicx}	
\usepackage{amsmath}	
\usepackage{subcaption}

\title[Eccentric Be/X-ray Binaries]{Decretion disc evolution and neutron star accretion in short-period eccentric Be/X-ray binaries}

\author[R. G. Rast et al.]{
R. G. Rast,$^{1}$\thanks{E-mail: krast@uwo.ca}
C. E. Jones,$^{1}$
M. W. Suffak$^{1}$
and A. C. Carciofi$^{2}$
\\
$^{1}$Department of Physics and Astronomy, Western University, London, ON N6A 3K7, Canada\\
$^{2}$Instituto de Astronomia, Geof\'ica e Ci\^encias Atmosf\'ericas, Universidade de S\~ao Paulo, S\~ao Paulo, SP 05508-090 Brazil\\
}

\date{Accepted XXX. Received YYY; in original form ZZZ}

\pubyear{\the\year{}}

\begin{document}
\label{firstpage}
\pagerange{\pageref{firstpage}--\pageref{lastpage}}
\maketitle

\begin{abstract}
We examine Be star discs in highly eccentric Be/X-ray systems. We use a three-dimensional smoothed particle hydrodynamics (\textsc{SPH}) code to model the structure of the Be star disc and investigate its interactions with the secondary star over time. We use system parameters consistent with the eccentric, short-period (P $\approx$ 16 d) Be/X-ray binary A0538-66 as the basis for our models. We explore a range of system geometries by incrementally varying the misalignment angle of the neutron star's orbital plane with respect to the primary star's equatorial plane to cover a complete range from coplanar prograde to coplanar retrograde. For all simulations, we follow the evolution of the disc’s total mass and angular momentum as well as the average eccentricity and inclination with respect to the equatorial planes of both the primary and secondary. We also determine the neutron star accretion rates. We find that the high eccentricity of the binary orbit causes all calculated disc parameters to vary with orbital phase in all models. The amplitude of these variations is negatively correlated with misalignment angle for models with misalignment angles less than 90$^{\circ}$, and positively correlated for models  with misalignment angles greater than 90$^{\circ}$. Accretion rates are affected by the number of particles the neutron star interacts with as well as the length of the interaction time between the particles and the neutron star. We find that accretion rates are largest for models with misalignment angles less than 90$^{\circ}$, and smaller for models with those greater than 90$^{\circ}$.

\end{abstract}




\section{Introduction}

For decades, Be stars have been defined as rapidly rotating stars of spectral type B that have been observed to show Balmer lines in emission  \citep{jas81, col87}. Among classical Be stars, these emission lines are produced by gaseous rotating discs that are formed and sustained by stellar mass loss. While this description of the Be star disc dates back nearly a century to \citet{str31}, who imagined a ``nebulous ring,'' the cause of the disc-building phenomenon has yet to be fully explained. To date, the leading hypotheses have involved some combination of rapid rotation and nonradial pulsations \citep{kri75, owo06, baa18, lab22, dod24}. 

Be star discs are dynamic, undergoing periods of growth and dissipation as the star’s mass loss rate changes. Observationally, these variations can be traced through changes in Balmer line emission strength and morphology, photometric brightness, as well as polarization degree and angle \citep{dra14, bar18, baa23}. Many Be stars have shown variations in these regimes with timescales of months to years. This variability over the span of human lifetimes makes Be stars compelling objects to observe and model. 

The most successful description of the Be star disc is provided by the viscous decretion disc (VDD) model developed by \citet{lee91}. In the VDD, the central star expels parcels of mass and angular momentum into the inner disc at its equator. These parcels are then carried outward from the star through viscous torques to create a radially extended disc. Optical long-base interferometry observations have shown that these discs are geometrically thin \citep{qui94} and polarimetry indicates they have opening angles of about 2.5$^{\circ}$ \citep{woo97}. Observations of the H$\alpha$ spectral line, continuum excess in the infrared, and intrinsic polarization all indicate that the disc material orbits according to Kepler's laws \citep{mar87, whe12, mar18}.

Many Be stars are known to exist in binary systems. For B-type stars, the binary fraction is estimated to be more than 50\% \citep{mar24a}. Since Be stars represent roughly one in five galactic B-type stars \citep{bod20}, we might expect a similar proportion. However, the exact percentage of Be stars with companions has not been established with certainty, in part due to a lack of large-sample surveys \citep{mar24a}. Their most frequently detected companion stars are subdwarf OB stars, stripped companions, and neutron stars \citep{coe15, wan21, elb21, mar24a}.

Be/X-ray binaries represent a class of Be binary systems with neutron star companions. Most of these neutron stars emit X-ray pulses \citep{rei11}. Be/X-ray binaries have wide orbits, with periods usually between 10 and 300 days, and moderate to high eccentricities \citep{oka07}. The core-collapse supernova event that creates the neutron star can produce a ``natal kick'' which accelerates the neutron star over timescales of 10 seconds or less \citep{lai06, col22, hir24}. This acceleration may eject the neutron star entirely from the system. If the binary system survives this event, the new path of the neutron star can become eccentric and misaligned with respect to the primary star's equatorial plane \citep{hug99, mar11, sal20, mar24b}, although the details have yet to be fully worked out \citep{mar09a}. For highly eccentric binary systems, the mean observed spin-orbit misalignment angle has been found to be larger than in systems with smaller eccentricities \citep{mar24b}. Be/X-ray binaries, therefore, provide opportunities to study the effects of eccentric, misaligned binary systems on the Be star disc.

Due to the moderate to high eccentricities of Be/X-ray binaries, the periastron distances can be quite small, on the order of a few times the radius of the primary star (R$_{\star}$). During closest approach, therefore, the neutron star's orbital path can intersect the disc, allowing it to gravitationally interact with disc material. In some cases, the neutron star can accumulate enough matter from the Be star disc to form its own accretion disc. As particles become unbound to the primary star (the Be star) and enter the Roche lobe of the secondary (the neutron star), some are unable to form stable orbits in an accretion disc and will fall directly onto the neutron star \citep{oka01}. The kinetic energy of this in-falling material is converted to electromagnetic energy in the form of X-ray radiation. In many cases, enhanced X-ray activity occurs once per orbital period (usually at periastron), and is classified as periodic ``Type I'' X-ray activity characterised by luminosities no greater than 10$^{37}$ erg s$^{-1}$ \citep{oka01, rei11, oka13}. However, non-periodic, episodic Type II outbursts with luminosities exceeding 10$^{37}$ erg s$^{-1}$ are also observed \citep{oka13}. 

Foundational modelling efforts on Be/X-ray binaries were conducted by \citet{oka02} and \citet{oka07}. These works explored the effects of the neutron star on the Be star disc, finding that the companion can produce two-waved spiral arms as well as truncate the Be star disc in all but the most eccentric cases. In the past decade, important advances have been made in modelling Be binary systems. \citet{suf22} explored disc growth and dissipation among equal-mass binaries in circular orbits with misalignment angles up to 60$^{\circ}$ and observed disc tearing in some configurations. \citet{suf23} used a radiative transfer code to predict the thermal structure and observed H$\alpha$ lines, $V$-band magnitude, and intrinsic polarization for tilted and warped Be star discs. \citet{mar14} modelled a moderately eccentric Be/X-ray system with large misalignment angles, focusing on how precession and warping of the disc may affect X-ray outbursts from the neutron star. \citet{pan16} studied coplanar binaries with a range of eccentricities, and also varied the direction of the disc’s rotation with respect to the binary’s orbital motion to create both prograde and retrograde orbits. \citet{cyr17} studied circular binaries with misalignment angles up to 60$^{\circ}$ and varying periods. \citet{ove24} modelled retrograde discs in Be/X-ray systems with circular binary orbits, with a focus on the effects of disc inclination. So far, no model-based study has presented the effects of a broad range of misalignment angles on eccentric Be binary systems.

In this paper, we present a range of \textsc{SPH} models for an eccentric Be/X-ray binary system. We incrementally vary the misalignment angle of the secondary star's orbital plane with respect to the disc from the coplanar prograde case, where the secondary star orbits in the same direction as the disc, to the coplanar retrograde case, where the secondary star's orbital path is antiparallel to the disc. We study the evolution of the disc in each of these cases as it builds and reaches a quasi-steady state. We measure the rates of accretion onto the neutron star in each case, and examine the orbital phases where this accretion is most efficient. This work lays the foundation for a future study on the observational properties of these systems. We begin with a description of our code and calculations in Section~\ref{sec:methods}, followed by our results in Section~\ref{sec:results} and our discussion in Section~\ref{sec:discussion}. Finally, our conclusions are provided in Section~\ref{sec:conclusion}.

\section{Methods}
\label{sec:methods}

\begin{table}
	\centering
	\caption{Parameters of the \textsc{SPH} models.}
	\label{tab:sys_params}
	\begin{tabular}{lccr}
		\hline
		Parameter & Value \\
		\hline
		Primary mass & 8.84 M$_{\odot}$ \\
		Secondary mass & 1.44 M$_{\odot}$ \\
		Primary radius & 10 R$_{\odot}$ \\
		Secondary radius & 10 km \\
            Primary T$_{\rm{eff}}$ & 25,000 K \\
            Orbital period & 16.6409 d \\
		Eccentricity & 0.72 \\
		$\alpha_{\rm{SS}}$ & 0.5 \\
            Injection radius & 1.04 R$_{\star}$ \\
            Mass loss rate & 1$\times$10$^{-8}$ M$_{\odot}$ yr$^{-1}$\\
            Misalignment angle & \makecell{0$^{\circ}$/30$^{\circ}$/45$^{\circ}$/60$^{\circ}$/75$^{\circ}$  \\ 105$^{\circ}$/120$^{\circ}$/135$^{\circ}$/150$^{\circ}$/180$^{\circ}$} \\
		\hline
	\end{tabular}
\end{table}

\begin{figure}
	\includegraphics[width=\columnwidth]{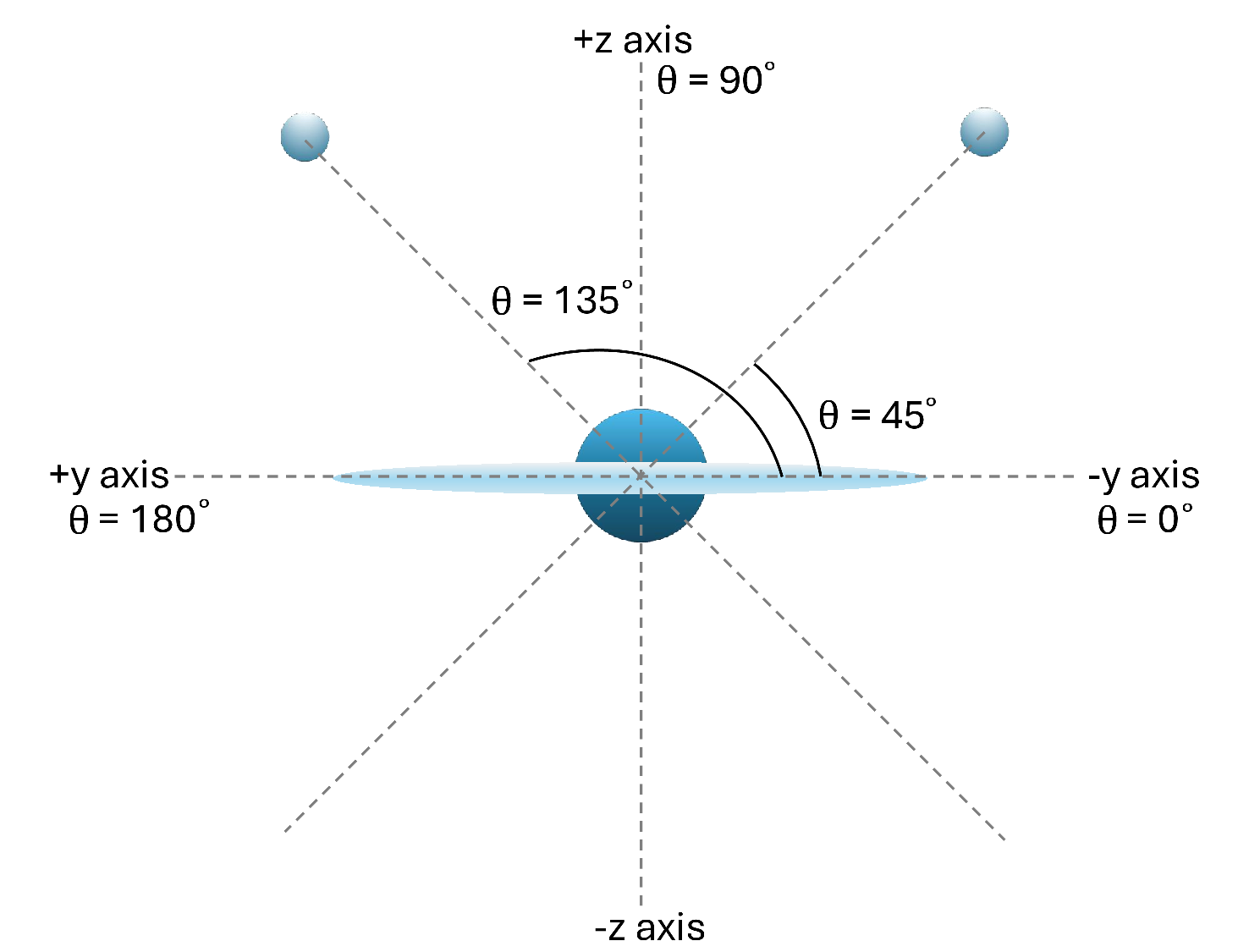}
    \caption{Schematic representing the coordinate system used for our misaligned binary systems. The primary star is at the center. Two configurations of the system are shown; one where the orbital plane of the secondary star (small blue circle) is misaligned by 45$^{\circ}$ with respect to the $x$-$y$ plane, and another where it is misaligned by 135$^{\circ}$.}
    \label{schematic}
\end{figure}

We use the 3D smoothed particle hydrodynamics (\textsc{SPH}) code developed by \citet{ben90a} and refined by \citet{bat95}. This code was further modified by \citet{oka02} for the purposes of studying decretion discs around Be stars. It has been used in studies such as \citet{pan16}, \citet{cyr17} and \citet{suf22} to simulate Be star discs in a variety of configurations. This code models the Be binary system as two sink particles with set masses representing the primary and secondary stars, and a collection of gas particles which make up the disc. We set the temperature of the gas particles to 60\% of the primary star's effective temperature \citep{mil99a, mil99b}. The mass of all injected particles is equal, determined by dividing the mass loss rate evenly among the total number of injected particles. In our simulations, each injected particle has a mass of 4.6 $\times$ 10$^{-15}$ M$_{\odot}$. Particles which lose enough angular momentum to fall within the radius of the primary star are accreted and deleted from the simulation. Particles that enter a variable accretion radius which surrounds the secondary star are also accreted. This accretion radius is defined as $0.05\,r_{\rm{L}}$ where $r_{\rm{L}}$ is the Roche lobe, approximated by
\begin{equation}
    r_{\rm{L}} \simeq 0.462 \left( \frac{q}{1+q} \right)^{1/3} D\,,
	\label{eq:roche_lobe}
\end{equation}
\noindent where $q$ is the ratio of the secondary stellar mass divided by the primary stellar mass, and $D$ is the distance between the two stars \citep{egg83}. 

In this work, we utilise particle splitting, an important modification to the code which increases the resolution in the outer disc \citep{rub25}. This method splits particles into 13 ``child'' particles, increasing the total number of particles and therefore the simulation's resolution, if they meet specific conditions. Particles may split if they are located at least 10 R$_{\star}$ from the primary, have less than 70 neighbour particles, have masses no smaller than 4$\times$10$^{-4}$ of the initial particle mass, and the square of the smoothing length is greater than 10\% the distance between the particle and the center of mass. Particles which enter the Roche lobe of the secondary star are also split. These conditions are the same as those used by \citet{kit02} and implemented by \citet{rub25}. This modification to the code increases the resolution in regions of the disc that would otherwise have a limited number of particles. This allows us to trace more carefully the behaviour of the disc near the neutron star and provide more accurate estimates for the rates at which particles enter the secondary's sink radius.

The simulations begin with no disc around the Be star. In the first time step, we inject a shell of 10,000 particles and continue to add 40,000 particles per subsequent time step of $1/2\pi$ orbital periods ($P_{\rm{orb}}$). This injection occurs over a prescribed zone consisting of a ring with its inner edge at a radius of 1 R$_{\star}$, and extending 0.04 R$_{\star}$ outwards. This injection radius was also used by \citet{cyr17}, \citet{suf22} and \citet{rub25}.  We hold the mass loss rate constant throughout our simulations, injecting particles with enough angular momentum to orbit at a distance of 1 R$_{\star}$ from the primary. A large fraction of the injected particles are quickly accreted by the primary star, with the rest drifting outwards to create the disc. The secondary star is initially located at apastron, so the disc has approximately half an orbital period to build before its first encounter with the secondary star. 

The radial motion of the particles in the disc is primarily governed by viscous torques. The disc shear viscosity, $\nu$, in the \textsc{SPH} code is described by the Shakura-Sunyaev prescription:
\begin{equation}
    \nu=\alpha_{\rm{SS}} c_s H\,,
	\label{eq:ss_viscosity}
\end{equation}
where $c_s$ is the speed of sound in the disc, and $H$ is the disc scale height \citep{sha73, oka02}. $\alpha_{\rm{SS}}$ is a dimensionless constant which takes a value between 0 and 1 \citep{rim18}. We set $\alpha_{\rm{SS}} = 0.5$ for all models. It is possible to describe an approximate relationship between $\alpha_{\rm{SS}}$ and the \textsc{SPH} artificial viscosity parameter $\alpha_{\rm{SPH}}$ through 
\begin{equation}
    \alpha_{\rm{SS}}=\frac{1}{10} \alpha_{\rm{SPH}} \frac{h}{H}\,,
	\label{eq:sph_viscosity}
\end{equation}
where $h$ represents the smoothing length of the particles \citep{oka02}.

We choose system parameters consistent with A0538-66, a Be/X-ray binary with a large eccentricity of $e = 0.72$ and a period of 16.6 days \citep{raj17}. For the primary star, we chose the effective temperature ($T_{\rm{eff}}$) and radius values used by \citealt{raj17}. For the companion, we choose the canonical neutron star mass of 1.44 M$_{\odot}$ and radius of 10 km \citep{oze16, cap20}. We varied the misalignment angle between the orbital plane of the neutron star to create a range of system configurations. The injected disc particles are assigned orbits in the $x$-$y$ plane, which is aligned with the equator of the Be star, so misaligned binary orbits are achieved by rotating the orbital plane of the secondary about the $x$-axis. In this system, misalignment angles of 0$^{\circ}$ and 180$^{\circ}$ represent coplanar prograde and coplanar retrograde binary orbits, respectively. Figure \ref{schematic} shows the coordinate system used to define our misalignment angles. The system parameters used in our simulations are summarised in  Table~\ref{tab:sys_params}.

This work studies the disc during its first 80 $P_{\rm{orb}}$. We calculate the specific angular momentum $j$ and inclination angle $i$ of each disc particle following equations (1) and (4) in \citet{suf22}. We find the eccentricity $e$ of the particles using
\begin{equation}
    e = \left\Vert\frac{\bm{v} \times \bm{j}}{\mu} - \frac{\bm{r}}{ \| \bm{r} \|} \right\Vert \,,
	\label{eq:eccentricity}
\end{equation}
{where $\bm{r}$ and $\bm{v}$ are the position and velocity vectors of the particle, respectively, and $\mu = G M_{\rm{primary}}$. We find the total disc angular momentum $L$ by multiplying the specific angular momentum of each particle by its mass, then finding the sum for all particles. We then average all particle eccentricities and inclinations to find the average values for the disc. The total mass of the disc is found as the sum of the masses of all gas particles. The longitude of the ascending node $\Omega$ of the disc is calculated using
\begin{equation}
    \Omega = \arctan \frac{N_x}{N_y}\,,
	\label{eq:Omega}
\end{equation}

\noindent where $N_x$ and $N_y$ are the $x$ and $y$ components of the average line of nodes, respectively.

\section{Results}
\label{sec:results}

\begin{table*}
	\centering
	\caption{Summary of findings for all \textsc{SPH} models. The values shown here represent the mean for each parameter over an orbital period once the quasi-steady state has been reached.}
	\label{tab:findings_summary}
	\begin{tabular}{c c c c c c c c c}
		\hline
		\makecell{Misalignment \\ Angle ($^{\circ}$)} & \makecell{Final \\ $N_\mathrm{SPH}$} & \makecell{$M_{\rm{total}}$ \\ (\SI{e-11}{} $M_{\odot}$)} & \makecell{$L_{\rm{total}}$ \\ (\SI{e42}{\gram\per\square \cm \per \second})} & $e$  & \makecell{$i$ w.r.t\\Primary ($^{\circ}$)} & \makecell{$i$ w.r.t\\Binary ($^{\circ}$)} & $\Omega$ ($^{\circ}$) & \makecell{$\dot{M}_{\rm{NS}}$ \\($M_{\odot}$ yr$^{-1}$)} \\
		\hline
        0 & 35500 & 5.1 & 4.1 & 0.27 & 0 & 0 & -- & \SI{3.0e-10}{} \\
        30 & 38000 & 5.5 & 4.5 & 0.28  & 8 & 22 & 204 & \SI{2.1e-10}{} \\ 
        45 & 39000 & 5.7 & 4.4 & 0.25 & 12 & 37 & 204 & \SI{1.0e-10}{}\\
        60 & 42500 & 6.2 & 4.7 & 0.20 & 13 & 54 & 211 & \SI{5.9e-11}{} \\
        75 & 47800 &  6.6 &  4.9 & 0.19 & 11 & 70 & 212 & \SI{2.7e-11}{} \\
        105 & 48900 & 7.0 & 4.9 & 0.21 & 6 & 60 & 159 & \SI{5.2e-12}{} \\
        120 & 50000 & 7.1 & 4.9 & 0.20 & 8 & 64 & 117 & \SI{2.2e-12}{} \\
        135 & 49200 & 6.9 & 4.7 & 0.16 & 11 & 66 & 102 & \SI{2.6e-12}{}\\
        150 & 45600 & 6.5 & 4.3 & 0.12 & 13 & 64 & 91 & \SI{4.7e-12}{}\\
        180 & 34000 & 5.2 & 3.3 & 0.14 & 0 & 0 & -- & \SI{1.6e-11}{}  \\ 
		\hline
	\end{tabular}
\end{table*}

In this section, we present the results of our ten simulations, each with a different misalignment angle. Table~\ref{tab:findings_summary} summarizes results for each simulation averaged over an orbital period, once the quasi-steady state is reached. This table includes the final number of particles in each simulation ($N_\mathrm{SPH}$) in addition to the disc mass and average angular momentum, eccentricity, inclinations with respect to the primary and secondary stars, the average longitude of the ascending node, and the average accretion rate of the neutron star ($\dot{M}_{\rm{NS}}$). We then construct a complete analysis of the simulations, grouped by misalignment angle. Section~\ref{subsec:prograde_results} includes the results for models whose misalignment angles are less than 90$^{\circ}$ (which we define as prograde), while Section~\ref{subsec:retrograde_results} shows the results for our simulations with misalignment angles larger than 90$^{\circ}$ (which we define as retrograde). For the sake of conciseness, we refer to the model with a misalignment angle of 45$^{\circ}$ as our ``45$^{\circ}$'' model, and similarly to reference other simulations. 

\subsection{Prograde Models}
\label{subsec:prograde_results}

Section~\ref{subsubsec:prograde_evolution} provides the our findings for the average disc quantities, while Section~\ref{subsubsec:prograde_accretion_rates} summarizes the results for the accretion rates.

\subsubsection{Prograde Disc Evolution}
\label{subsubsec:prograde_evolution}

\begin{figure*}
	\includegraphics[width=\textwidth]{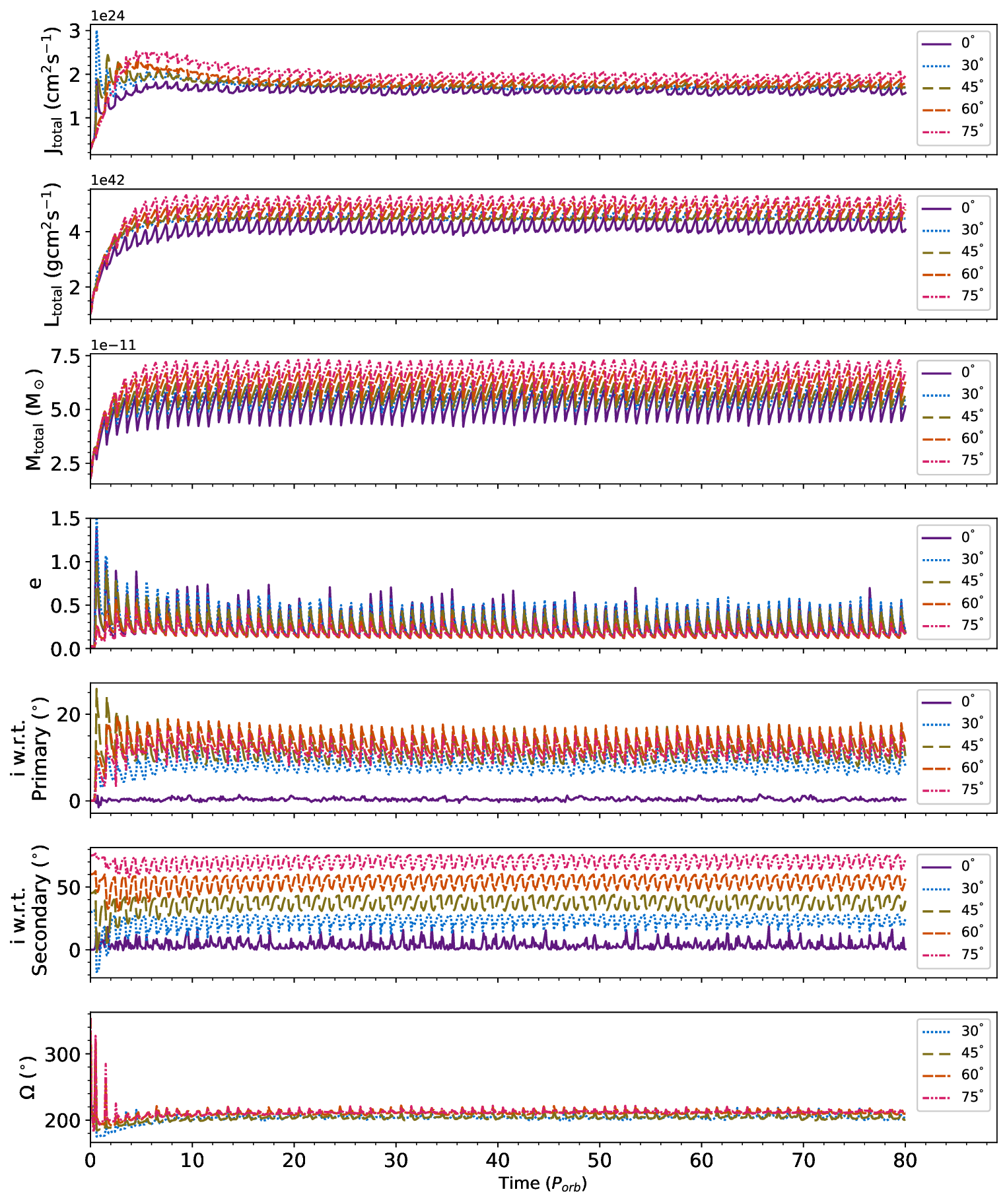}
    \caption{Evolution of the prograde models. Top to bottom: total disc specific angular momentum, total angular momentum, total disc mass, average disc eccentricity, average disc inclination relative to the primary and secondary stars, and average longitude of the ascending node for the disc particles. We omit the 0$^{\circ}$ model when plotting the longitude of the ascending node, as this parameter shows no trends for this misalignment angle. All quantities are plotted with respect to time, measured in $P_{\rm{orb}}$. The misalignment angle of each model is indicated in the legend on each panel.} 
    \label{fig:pro_evolution}
\end{figure*}

\begin{figure}
	\includegraphics[width=\columnwidth]{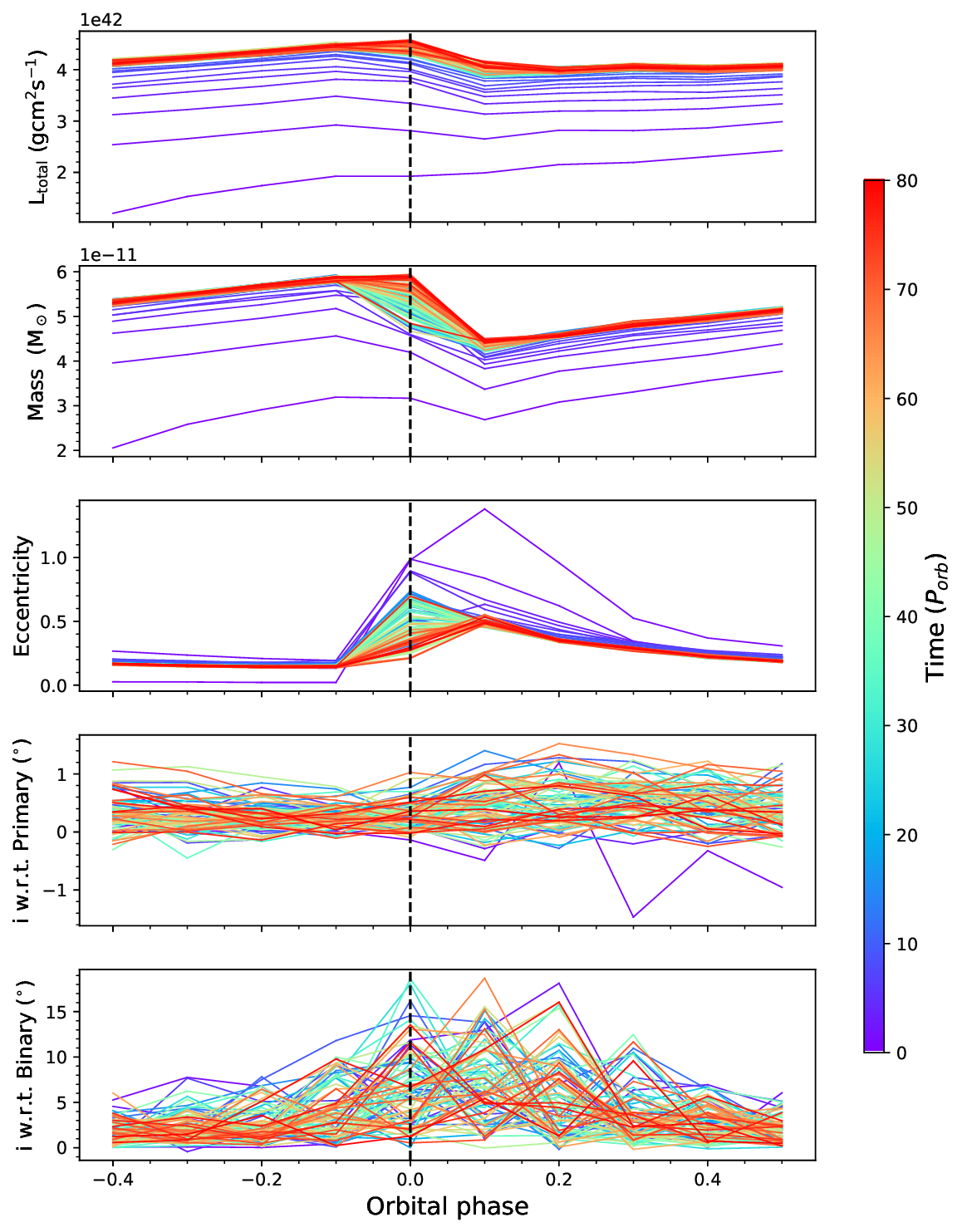}
    \caption{Top to bottom: total disc angular momentum, total disc mass, average disc eccentricity, and average disc inclination with respect to the primary and secondary stars, for the coplanar prograde simulation. All quantities are plotted as functions of orbital phase. The vertical dashed line represents the periastron.} 
    \label{fig:pro_coplanar_phased}
\end{figure}

\begin{figure}
	\includegraphics[width=\columnwidth]{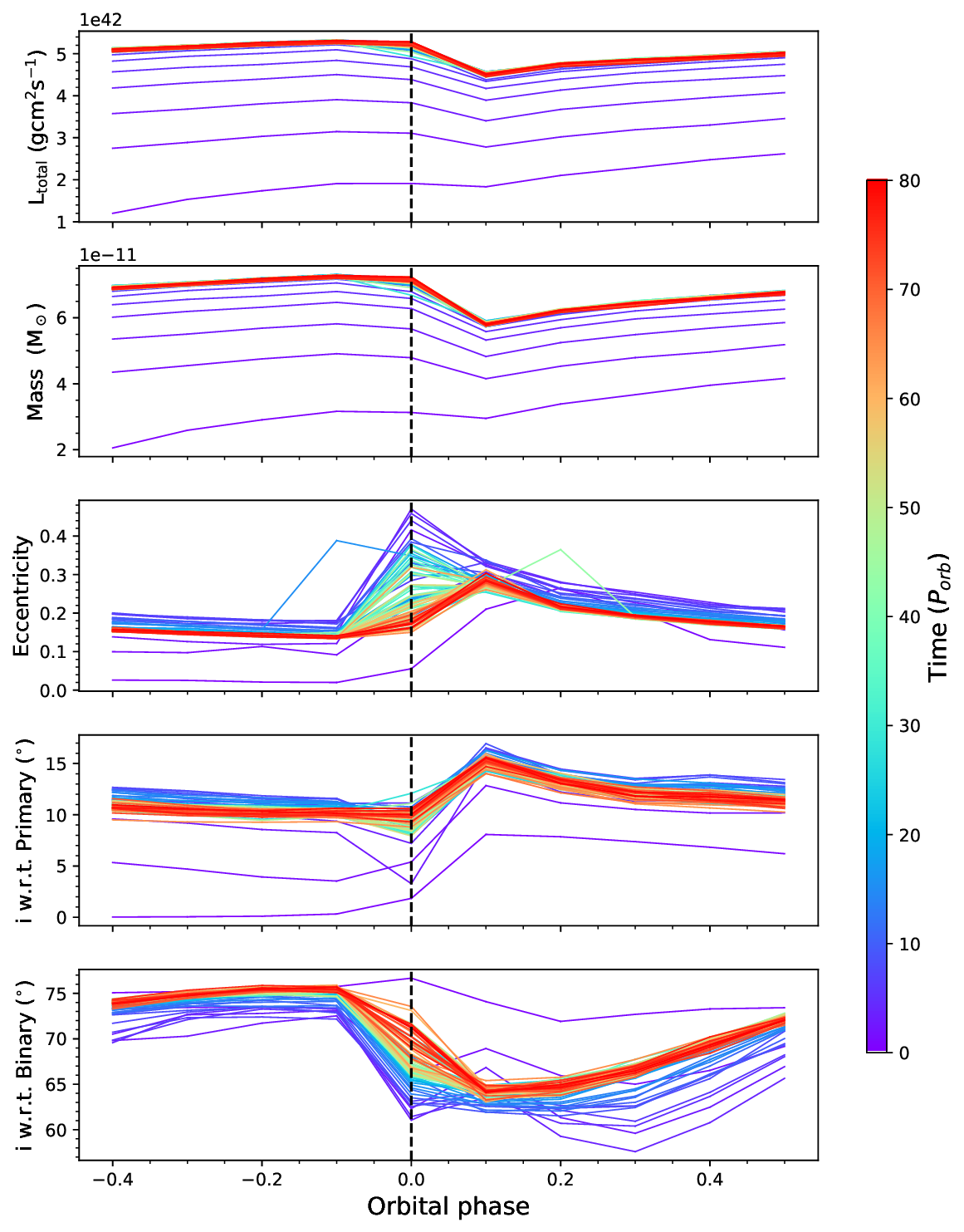}
    \caption{Same as Fig.~\ref{fig:pro_coplanar_phased}, but for the 75$^{\circ}$ simulation.} 
    \label{fig:pro_75_phased}
\end{figure}

\begin{figure*}
\centering
    \begin{subfigure}{\linewidth}
        \includegraphics[width=\linewidth]{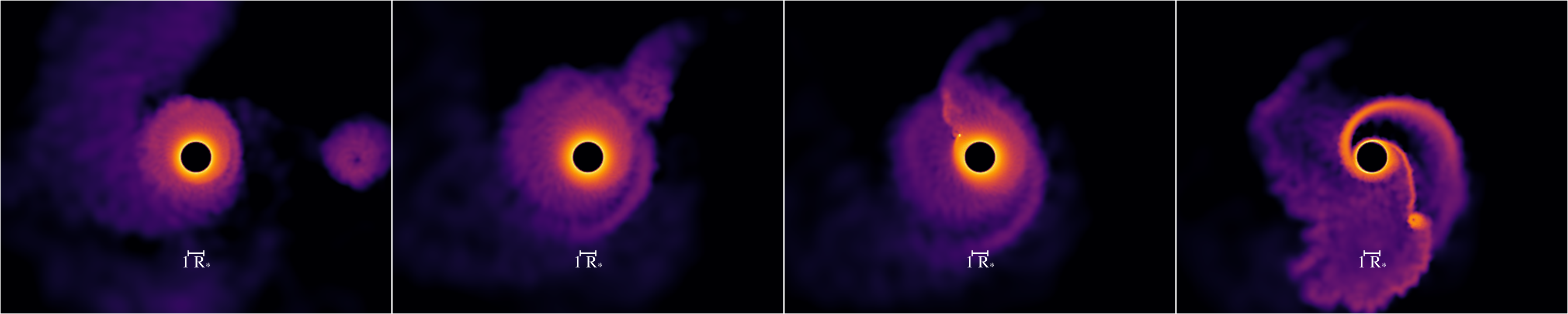}
    \caption{Coplanar.}
    \end{subfigure}
\hfil
    \begin{subfigure}{\linewidth}
        \includegraphics[width=\linewidth]{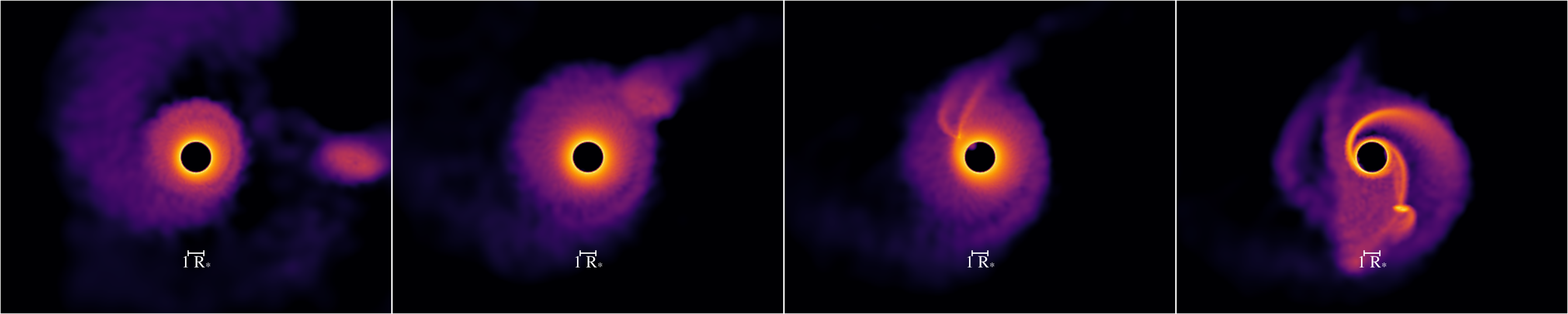}
    \caption{Misaligned by 30$^{\circ}$.}
    \end{subfigure}

    \begin{subfigure}{\linewidth}
        \includegraphics[width=\linewidth]{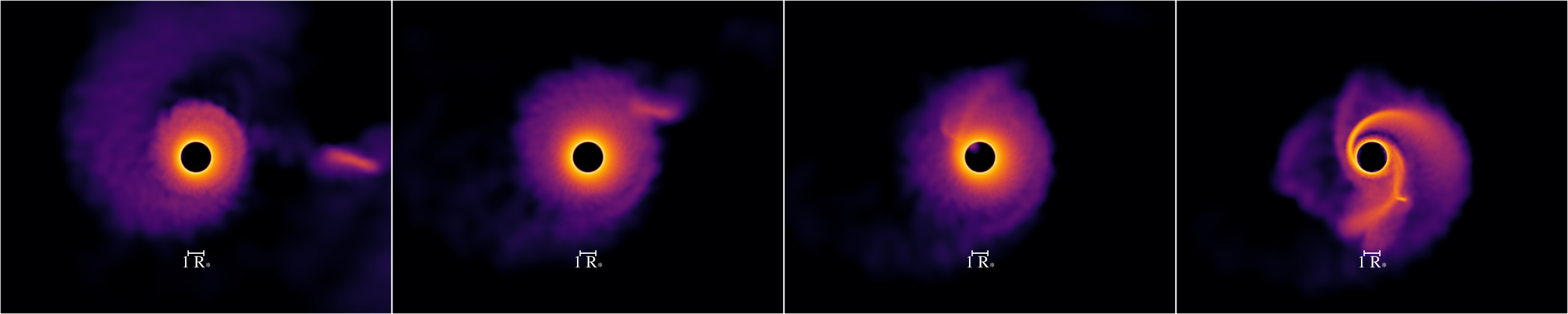}
    \caption{Misaligned by 45$^{\circ}$.}
    \end{subfigure}
\hfil
    \begin{subfigure}{\linewidth}
        \includegraphics[width=\linewidth]{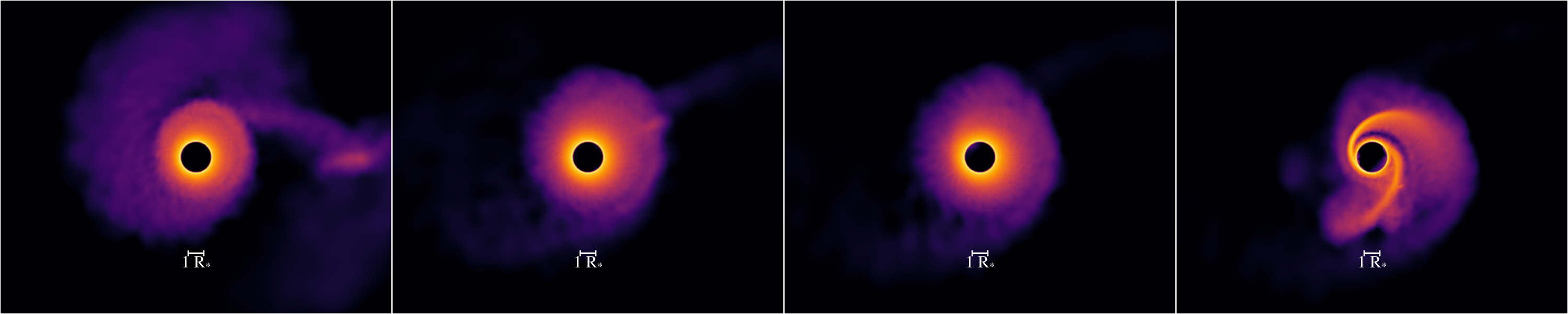}
    \caption{Misaligned by 60$^{\circ}$.}
    \end{subfigure}

    \begin{subfigure}{\linewidth}
        \includegraphics[width=\linewidth]{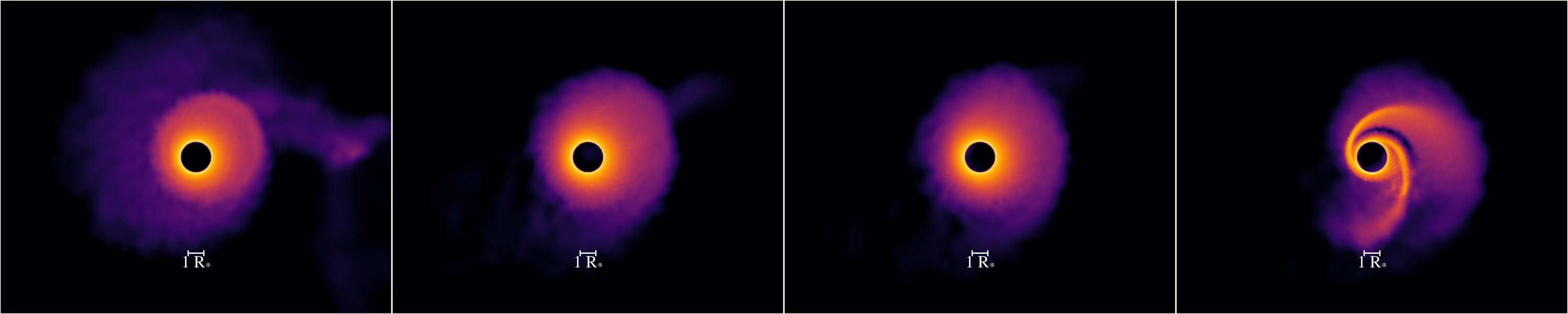}
    \caption{Misaligned by 75$^{\circ}$.}
    \end{subfigure}
\caption{Top-down view ($x$-$y$ plane) of the prograde simulations. From left to right, snapshots are taken at 60.0 $P_{\rm{orb}}$, 60.4 $P_{\rm{orb}}$, 60.5 $P_{\rm{orb}}$, and 60.6 $P_{\rm{orb}}$. In this view, the motion of the secondary star is counter-clockwise. Images rendered using \textsc{SPLASH} \citep{pri07}.}
    \label{fig:prograde_splash}
    \end{figure*}

\begin{figure*}
\centering
    \begin{subfigure}{\linewidth}
        \includegraphics[width=\linewidth]{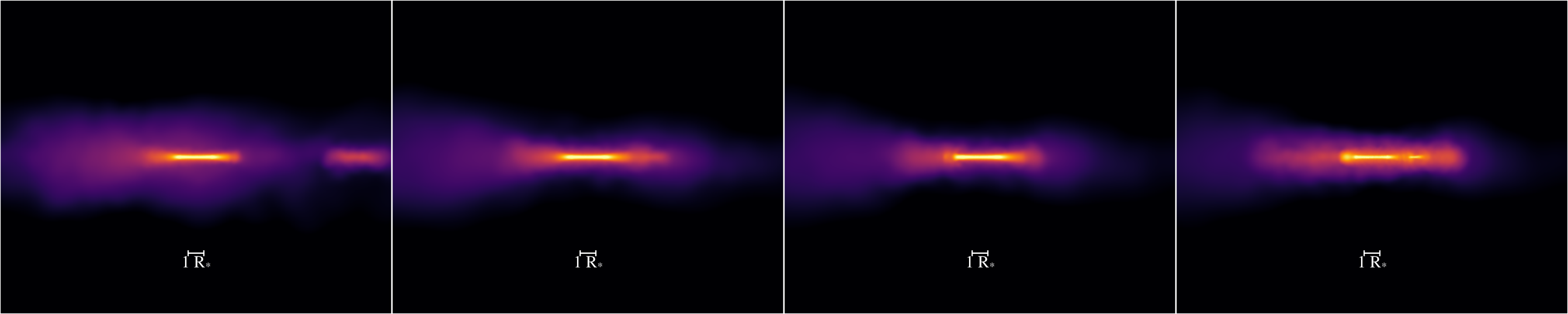}
    \caption{Coplanar.}
    \end{subfigure}
\hfil
    \begin{subfigure}{\linewidth}
        \includegraphics[width=\linewidth]{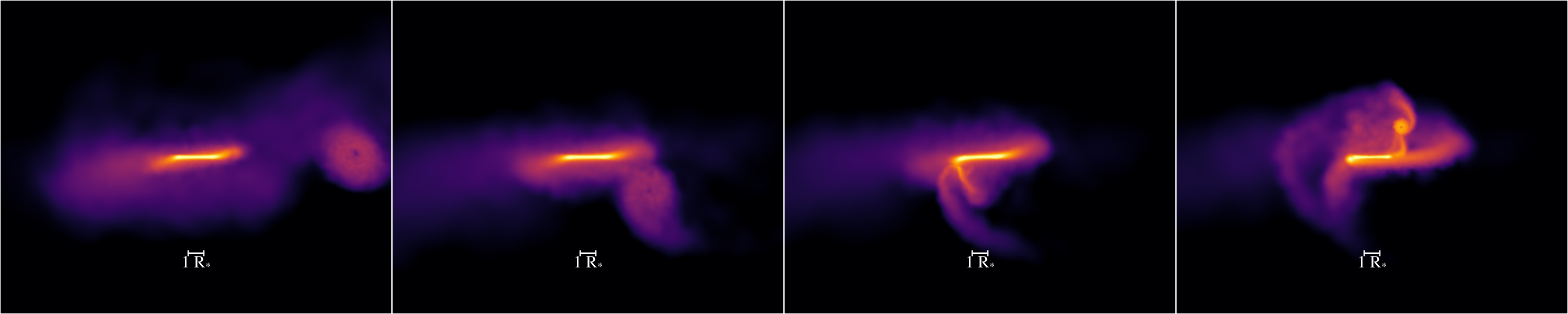}
    \caption{Misaligned by 30$^{\circ}$.}
    \end{subfigure}

    \begin{subfigure}{\linewidth}
        \includegraphics[width=\linewidth]{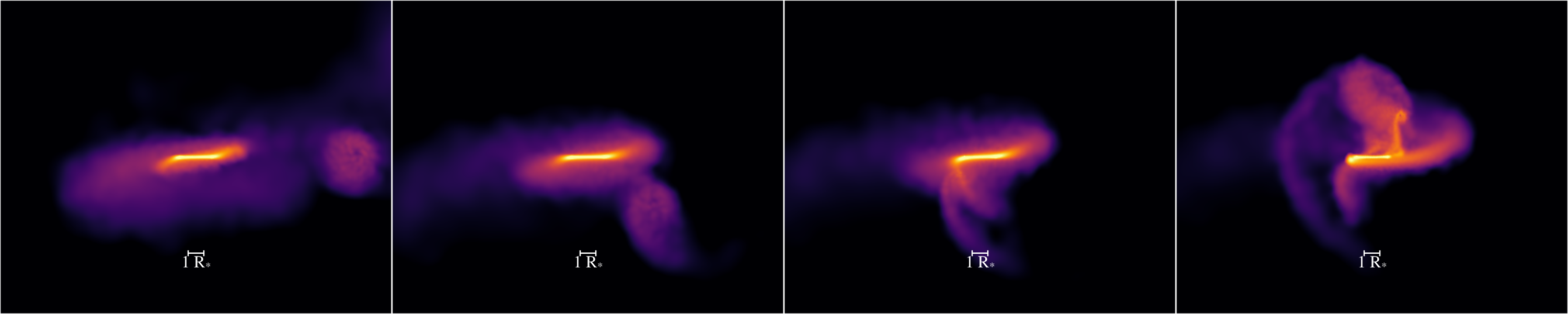}
    \caption{Misaligned by 45$^{\circ}$.}
    \end{subfigure}
\hfil
    \begin{subfigure}{\linewidth}
        \includegraphics[width=\linewidth]{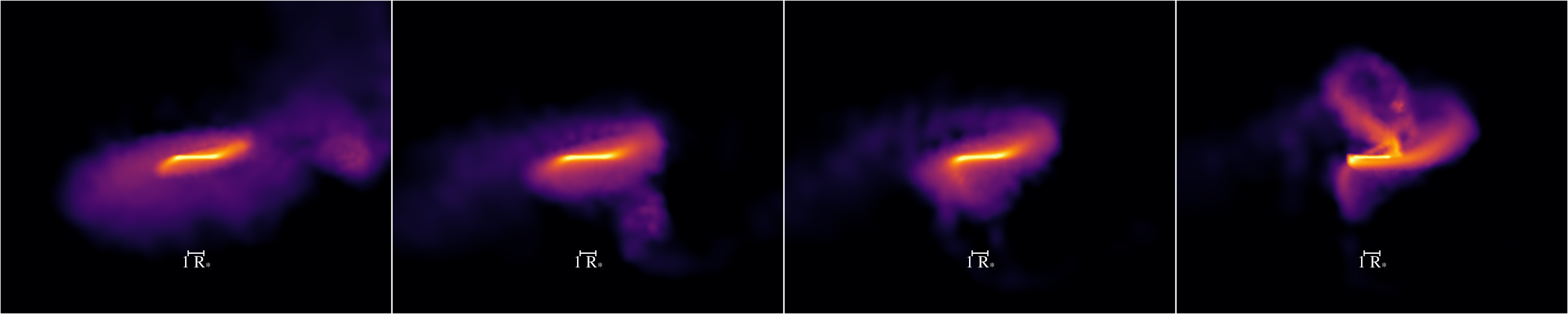}
    \caption{Misaligned by 60$^{\circ}$.}
    \end{subfigure}

    \begin{subfigure}{\linewidth}
        \includegraphics[width=\linewidth]{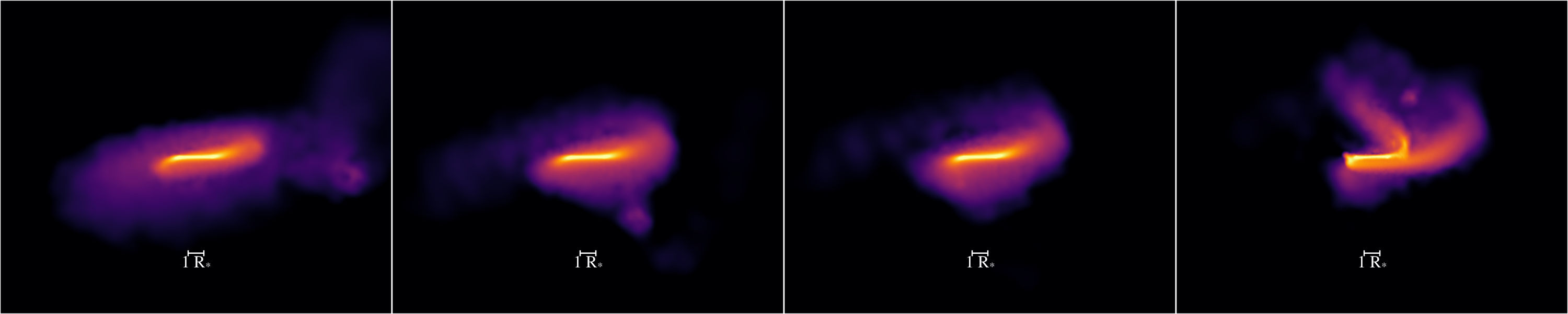}
    \caption{Misaligned by 75$^{\circ}$.}
    \end{subfigure}
\caption{Same as Fig.~\ref{fig:prograde_splash}, but for the prograde simulations in the $x$-$z$ plane. In this view, the motion of the secondary star is clockwise. Images rendered using \textsc{SPLASH} \citep{pri07}.}
    \label{fig:prograde_splash_xz}
    \end{figure*}

Figure~\ref{fig:pro_evolution} shows the evolution of the prograde discs. Here, we plot the total specific angular momentum, total angular momentum, eccentricity, inclination of the disc with respect to the primary's equatorial plane and with respect to the secondary's orbital plane, as well as the average longitude of the ascending node for the disc. Due to the eccentricity of the binary orbit, all computed quantities vary with orbital phase. The average eccentricity and inclination of the disc particles show significant disc disruption within the first few $P_{\rm{orb}}$, but the disc reaches a quasi-steady state shortly after. Misaligned discs show larger total disc masses and angular momenta than the coplanar system, but smaller average disc eccentricities. For simulations with low misalignment angles, eccentricity also varies cyclically over six $P_{\rm{orb}}$.
Misaligned binary orbits cause the disc to tilt away from the primary's equatorial plane, as seen in the fifth panel of Fig.~\ref{fig:pro_evolution}. Most particles do not align with the secondary's orbital plane, as demonstrated in the sixth panel of Fig.~\ref{fig:pro_evolution}. 

Since the disc properties fluctuate predictably within a given range once the quasi-steady state is reached, it is interesting to consider how the quantities shown in Fig.~\ref{fig:pro_evolution} change over the span of an orbital period. Fig.~\ref{fig:pro_coplanar_phased} shows the total disc angular momentum, mass, average eccentricity, and average inclinations with respect to the primary and secondary stars as a function orbital phase for the coplanar prograde simulation. Here, we define an orbital phase of 0 at periastron, consistent with \citet{mor11}. Figure~\ref{fig:pro_75_phased} shows the same quantities for the 75$^{\circ}$ simulation. We show these two plots as representative of the simulations with low misalignment angles. The data for the remaining prograde simulations are included in the Appendix. 

The quantities plotted in Fig.~\ref{fig:pro_coplanar_phased} all indicate a violent disruption throughout the disc near periastron. The disc's eccentricity and inclination relative to the primary and secondary begin responding to the secondary star's approach shortly before periastron, deviating from their baseline levels at phase -0.1. The total mass and angular momentum of the disc drop at periastron, reaching their minima 0.1 $P_{\rm{orb}}$ afterward. This orbital phase corresponds to the maximum eccentricity and inclinations. Subsequently, the disc recovers until its next encounter with the secondary star. During this recovery time, the disc mass and angular momentum rebuild due to the constant mass injection, while the eccentricity and inclination return to their baseline values. For the 75$^{\circ}$ model in Fig.~\ref{fig:pro_75_phased}, we see the same trend with time, but in this case the disc becomes more inclined with respect to the primary. 

We note that Figs.~\ref{fig:pro_coplanar_phased} and~\ref{fig:pro_75_phased} show quantities that are averaged over all particles. To investigate whether the velocities of the particles bound to the secondary, or escaping the system, may skew the average values over the whole disc at certain time steps, we compared the total mass of the particles bound to the primary star to those that are unbound  for the coplanar prograde case. We define the particles bound to the primary star to have negative specific energies with respect to it. Therefore, the remaining particles are either bound to the secondary star, or in escape orbits. We found that the majority of the mass is bound to the primary star, with the number of bound particles outnumbering the unbound by at least two orders of magnitude. Therefore, we do not expect the averaged disc properties to be impacted significantly by the velocities of the unbound particles, except in the first few orbits while the disc is being built and the relative number of particles unbound to the primary is larger.

To help visualise the disc's behaviour over an orbital period, Fig.~\ref{fig:prograde_splash} shows snapshots of the prograde discs in the $x$-$y$ plane at four key orbital phases. We selected orbital phases after 60 $P_{\rm{orb}}$ to represent the quasi-steady state behaviour of the discs. The left-most image in all panels shows the disc with the neutron star at apastron. Moving to the right, we see the disc 0.1 $P_{\rm{orb}}$ before periastron, then at periastron, and finally 0.1 $P_{\rm{orb}}$ after periastron. Two-armed spiral density enhancements are created in all prograde models, most predominantly in the coplanar case. The amplitude of the variations in the parameters shown in Fig.~\ref{fig:pro_evolution} can be used as tracers for the strength of these spiral arms; larger amplitudes are correlated with more defined spiral arms. Figure~\ref{fig:prograde_splash_xz} shows snapshots at the same orbital phases used in Fig.~\ref{fig:prograde_splash}, but in the $x$-$z$ plane. We see accretion streams connecting the primary and secondary discs in the misaligned cases during the time immediately following periastron, when the spiral arm density patterns are strongest. 

\subsubsection{Prograde Accretion Rates}
\label{subsubsec:prograde_accretion_rates}

\begin{figure*}
	\includegraphics[width=\textwidth]{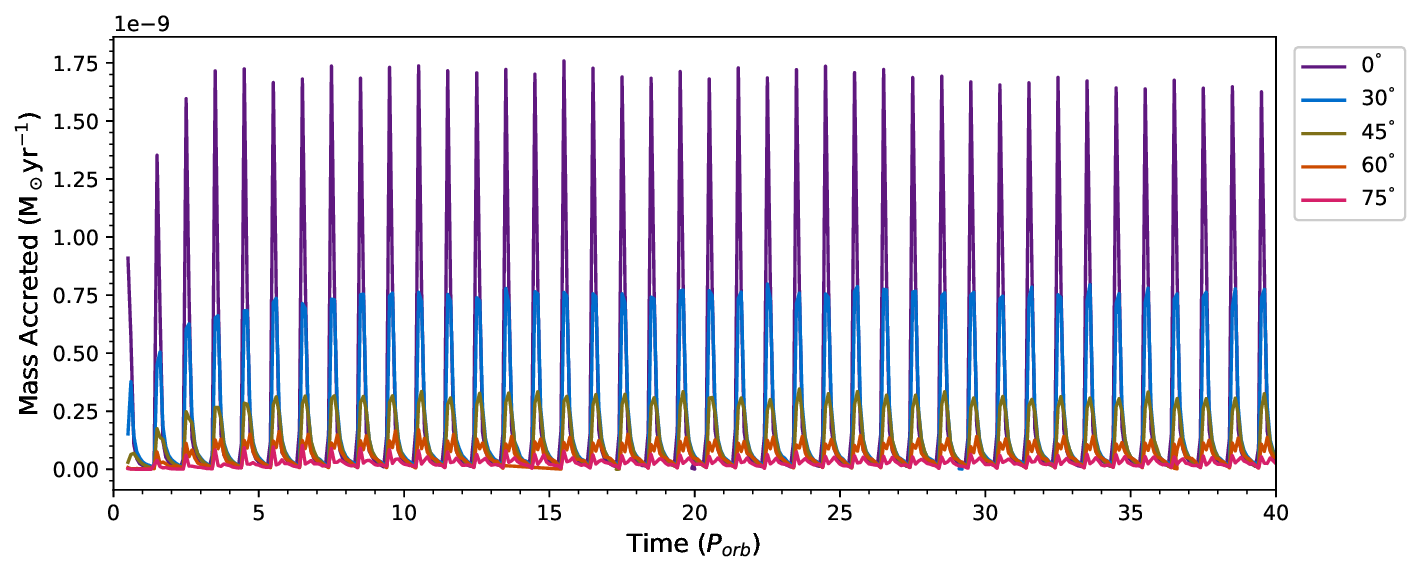}
    \caption{Accretion rates onto the secondary star, plotted as a function of time measured in $P_{\rm{orb}}$, for the prograde simulations. To better illustrate the dependence of accretion rates on orbital phase, we show the first 40 $P_{\rm{orb}}$. The misalignment angle of each model is indicated in the legend.} 
    \label{fig:pro_accrete_time}
\end{figure*}

\begin{figure}
	\includegraphics[width=\columnwidth]{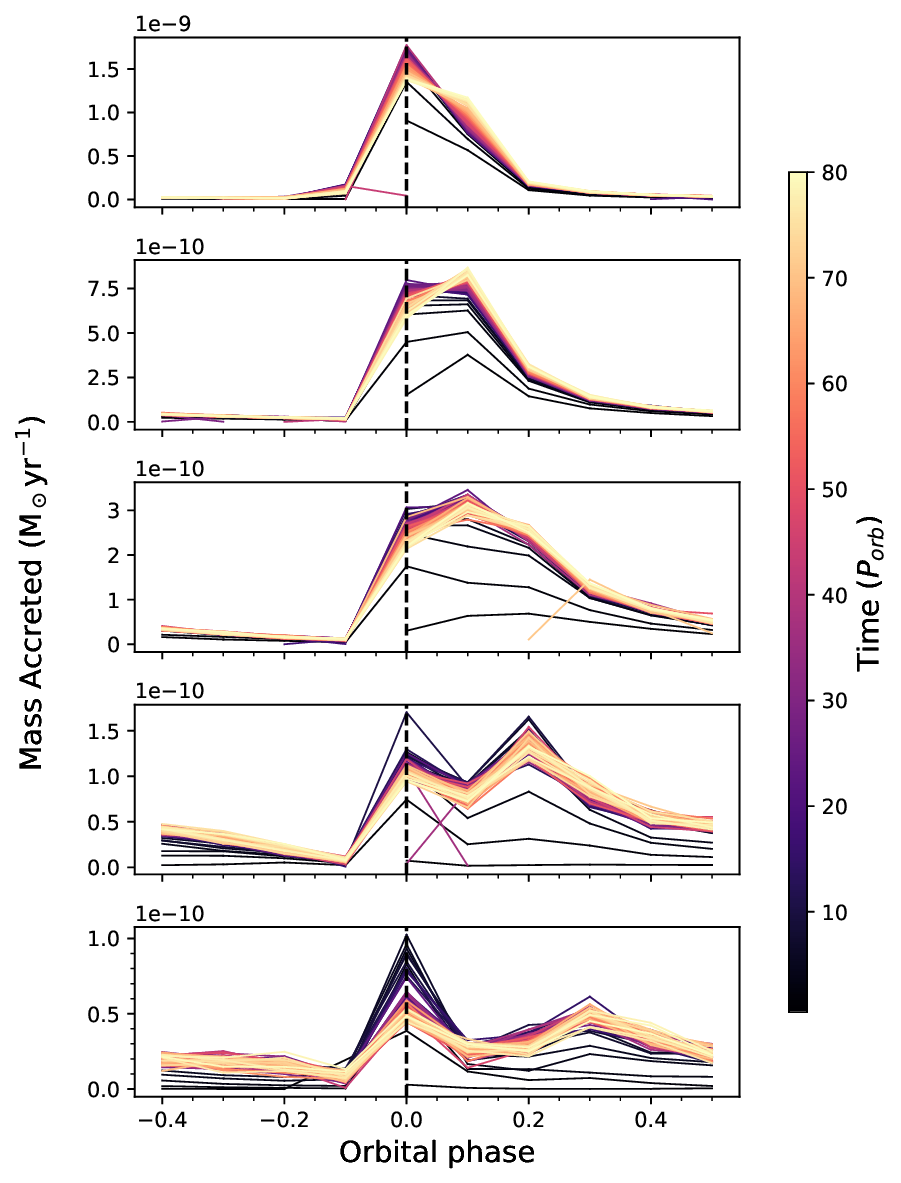}
    \caption{Accretion rates onto the secondary star as a function of time measured in $P_{\rm{orb}}$ since apastron, for the prograde simulations. The vertical dashed line represents the phase of periastron. Top to bottom: misalignment angles of 0$^{\circ}$, 30$^{\circ}$, 45$^{\circ}$, 60$^{\circ}$, and 75$^{\circ}$, respectively.} 
    \label{fig:pro_accrete_phased}
\end{figure}

To estimate the rate at which matter is accreted by the neutron star in each simulation, we find the rate at which matter enters the accretion radius of the secondary star, as defined in Section~\ref{sec:methods}. These values are plotted for the prograde simulations in Fig.~\ref{fig:pro_accrete_time}, for the first 40 $P_{\rm{orb}}$. Accretion rates are strongly correlated with orbital phase, and exhibit a negative correlation with misalignment angle. In addition to accumulating less matter, the neutron star in highly misaligned orbits also accretes the material at different orbital phases compared to models with misalignment angles $\leq$ 45$^{\circ}$. Rather than having a single, high accretion point at periastron, these models show multiple events of enhanced accretion per orbital period. This effect is clear in Fig.~\ref{fig:pro_accrete_phased}, which shows the accretion rate as a function of orbital phase for all prograde misalignment angles. While the peak accretion rate occurs at periastron for the coplanar case, it occurs later for slightly misaligned orbits and splits into two peaks for highly misaligned cases.

\subsection{Retrograde Models}
\label{subsec:retrograde_results}

\subsubsection{Retrograde Disc Evolution}
\label{subsubsec:retrograde_evolution}

\begin{figure*}
	\includegraphics[width=\textwidth]{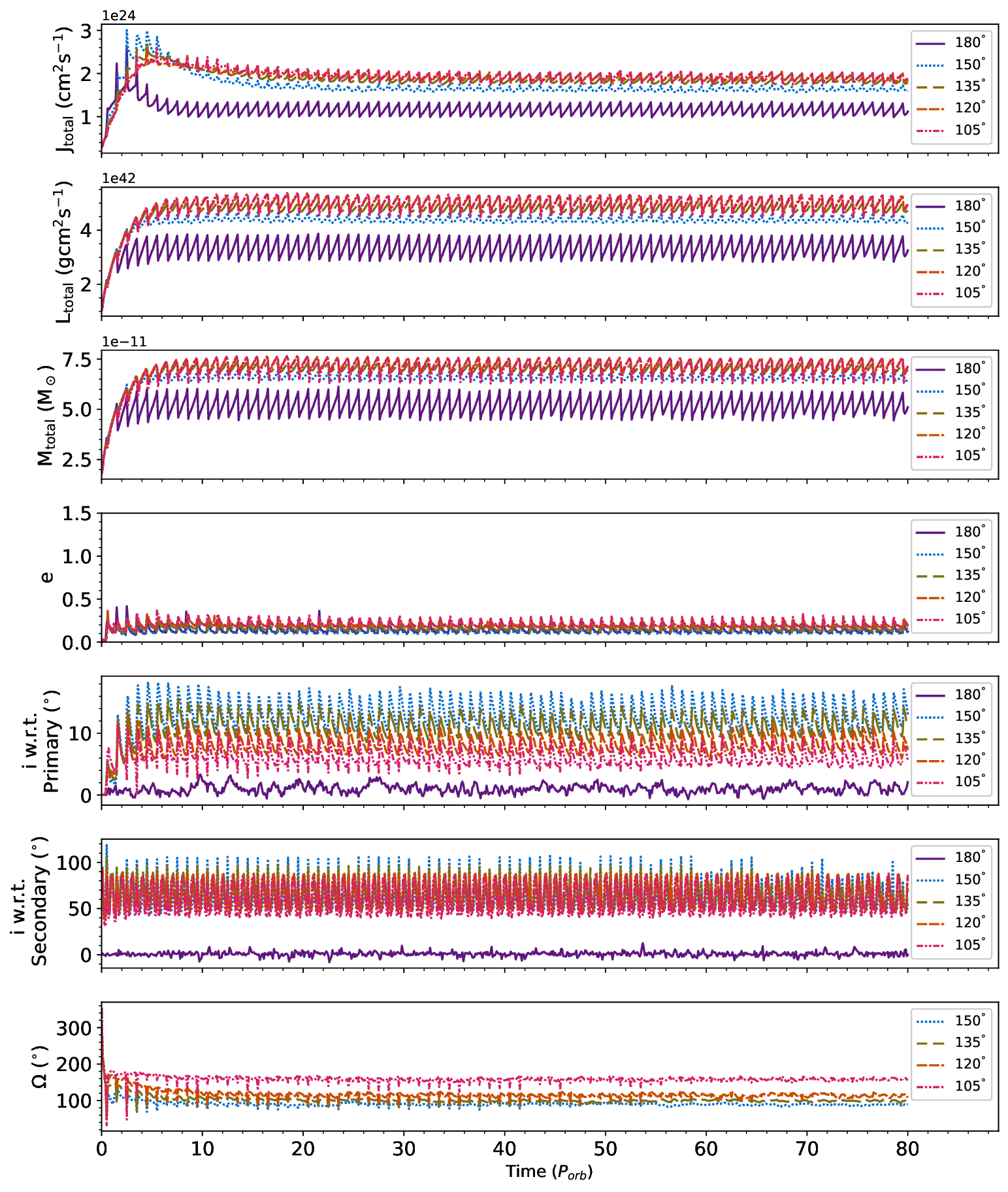}
    \caption{Same as Fig.~\ref{fig:pro_evolution}, but for the retrograde models. Note that the scales on the $y$-axes for the first five panels are the same as in Fig.~\ref{fig:pro_evolution}. In the sixth panel, the $y$-axis scale has been adjusted for these retrograde orbits. We omit the 180$^{\circ}$ model when plotting the longitude of the ascending node, as this parameter shows no trends for this misalignment angle.} 
    \label{fig:retro_evolution}
\end{figure*}

\begin{figure}
	\includegraphics[width=\columnwidth]{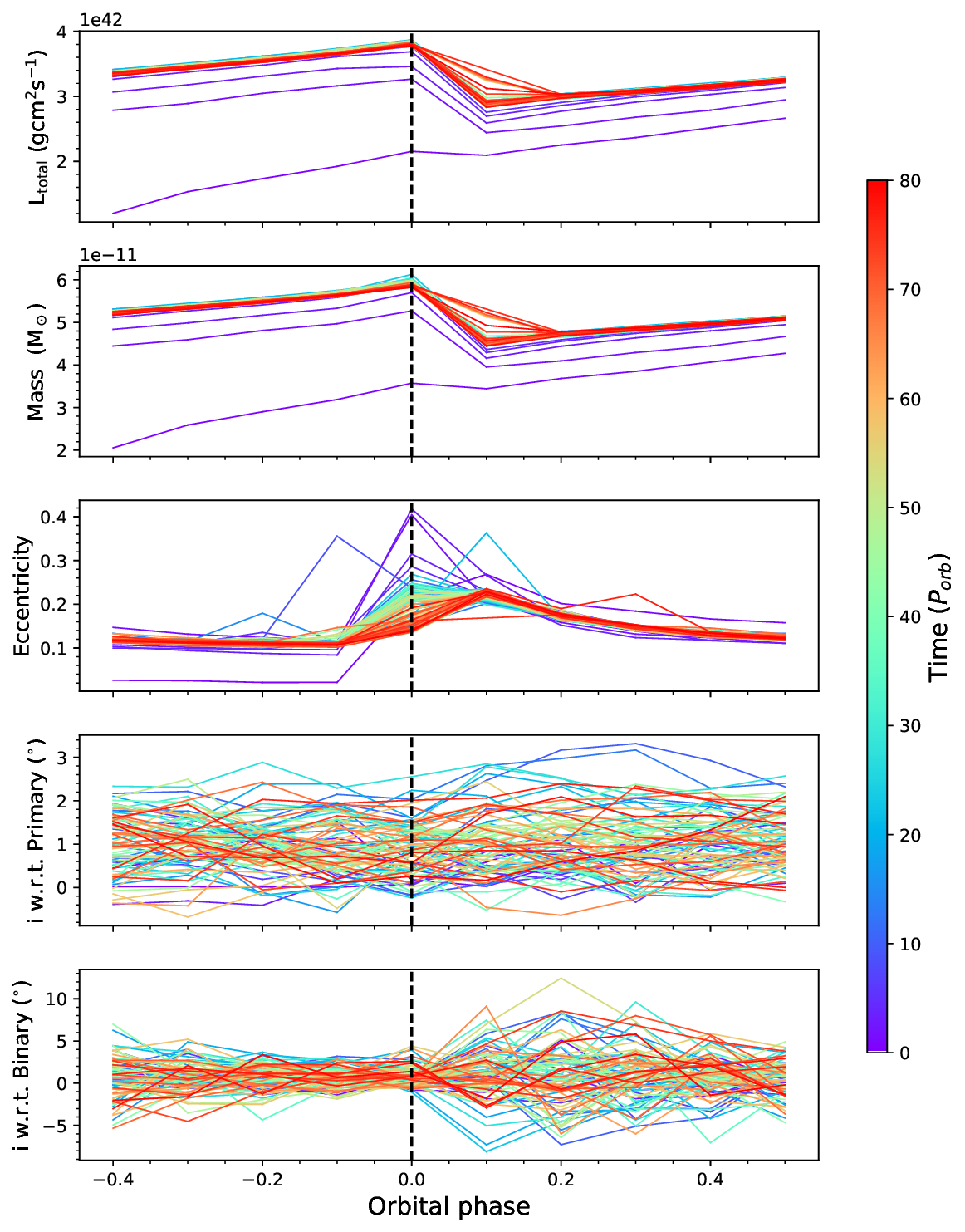}
    \caption{Same as Fig.~\ref{fig:pro_coplanar_phased}, but for the coplanar retrograde simulation.} 
    \label{fig:retro_coplanar_phased}
\end{figure}

\begin{figure}
	\includegraphics[width=\columnwidth]{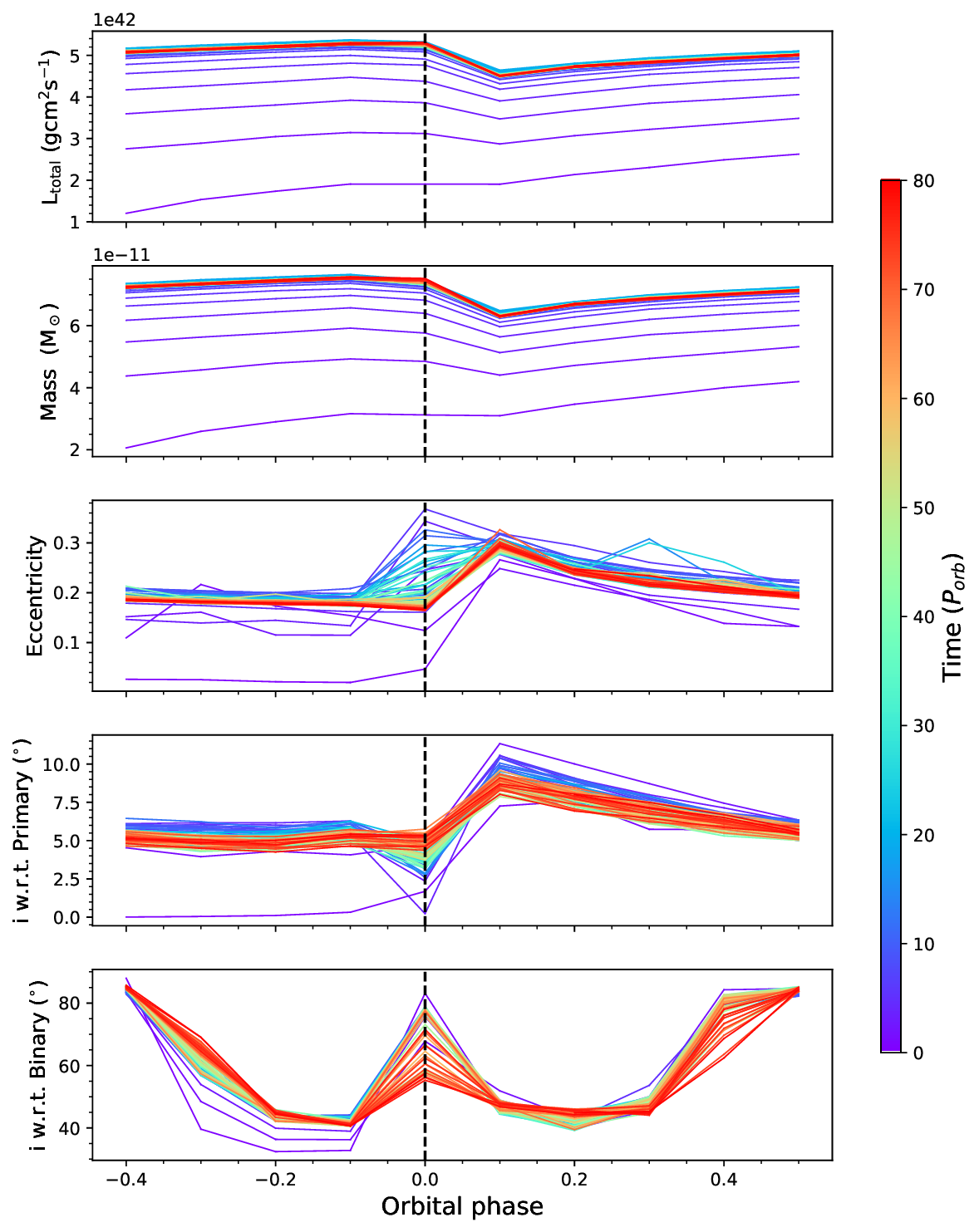}
    \caption{Same as Fig.~\ref{fig:pro_coplanar_phased}, but for the 105$^{\circ}$ simulation.} 
    \label{fig:retro_75_phased}
\end{figure}

\begin{figure*}
\centering
    \begin{subfigure}{\linewidth}
        \includegraphics[width=\linewidth]{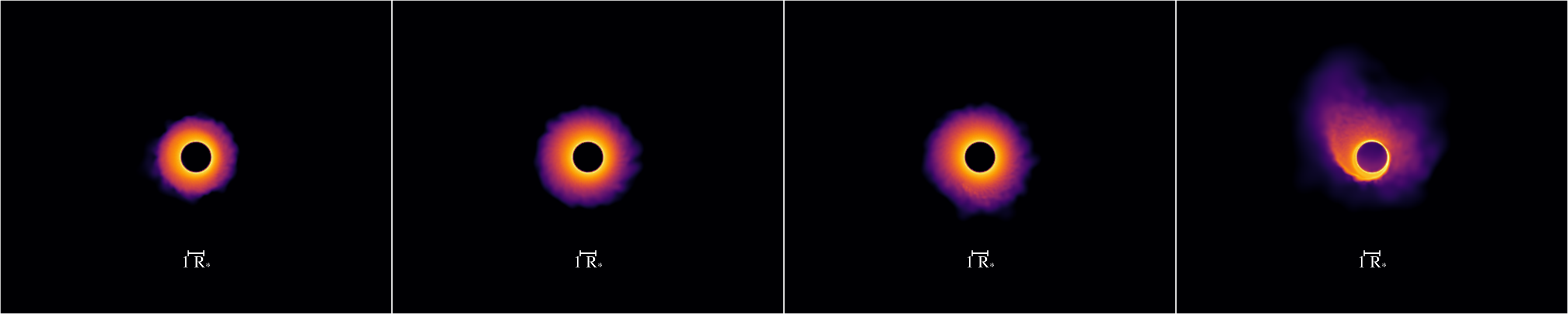}
    \caption{Misaligned by 180$^{\circ}$.}
    \end{subfigure}
\hfil
    \begin{subfigure}{\linewidth}
        \includegraphics[width=\linewidth]{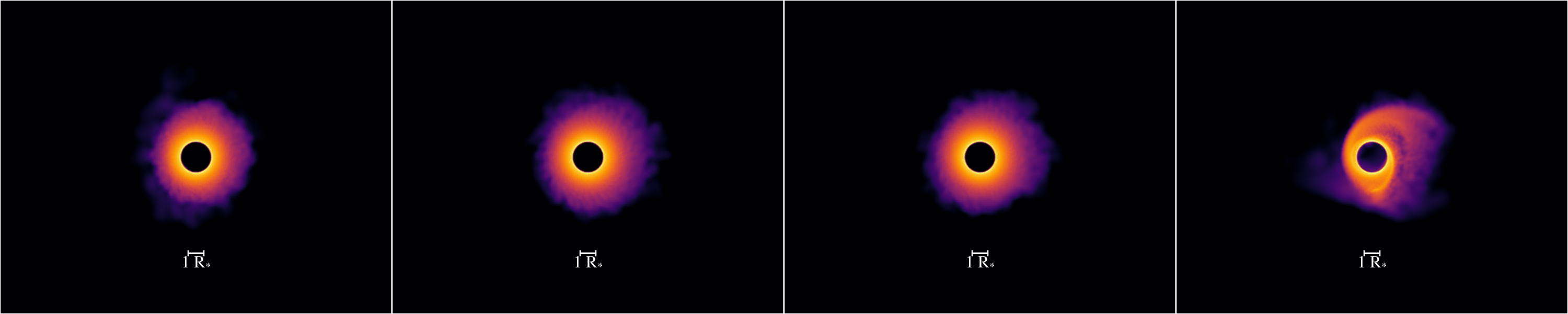}
    \caption{Misaligned by 150$^{\circ}$.}
    \end{subfigure}

    \begin{subfigure}{\linewidth}
        \includegraphics[width=\linewidth]{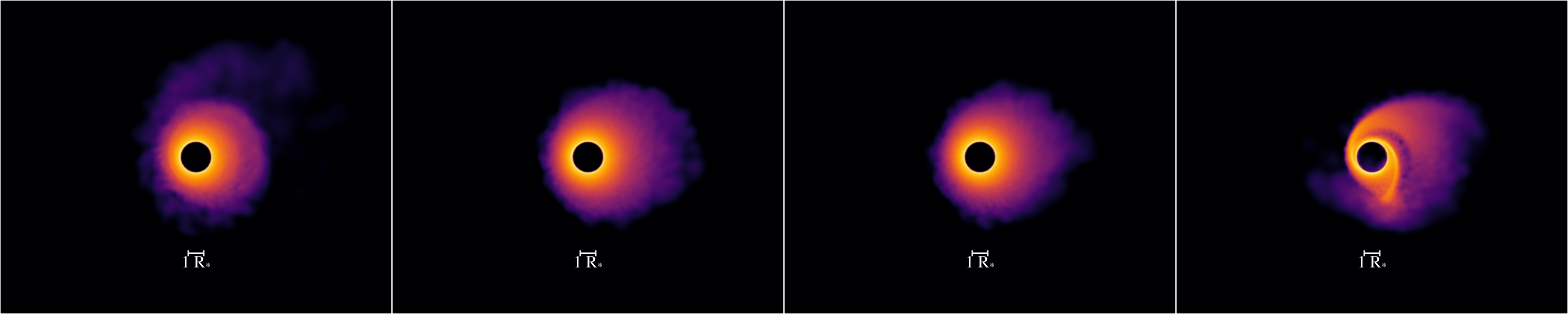}
    \caption{Misaligned by 135$^{\circ}$.}
    \end{subfigure}
\hfil
    \begin{subfigure}{\linewidth}
        \includegraphics[width=\linewidth]{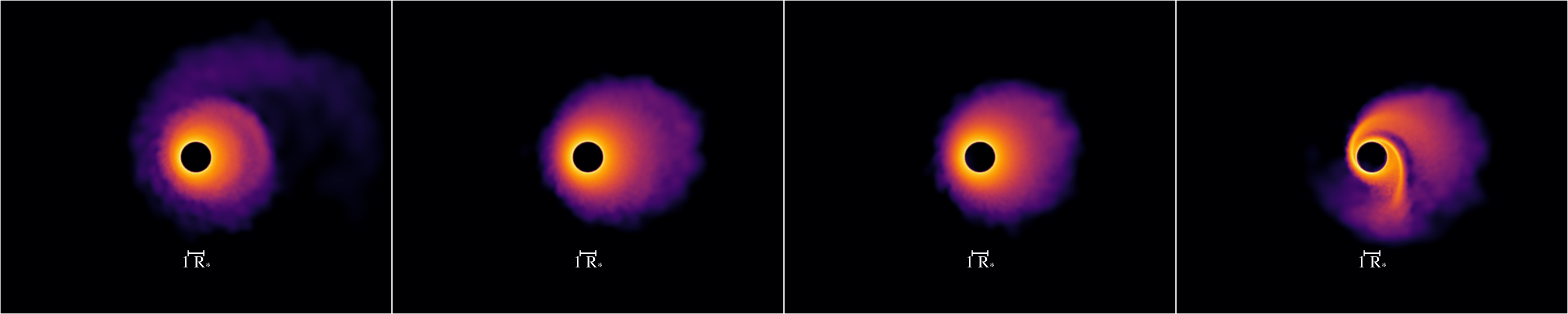}
    \caption{Misaligned by 120$^{\circ}$.}
    \end{subfigure}

    \begin{subfigure}{\linewidth}
        \includegraphics[width=\linewidth]{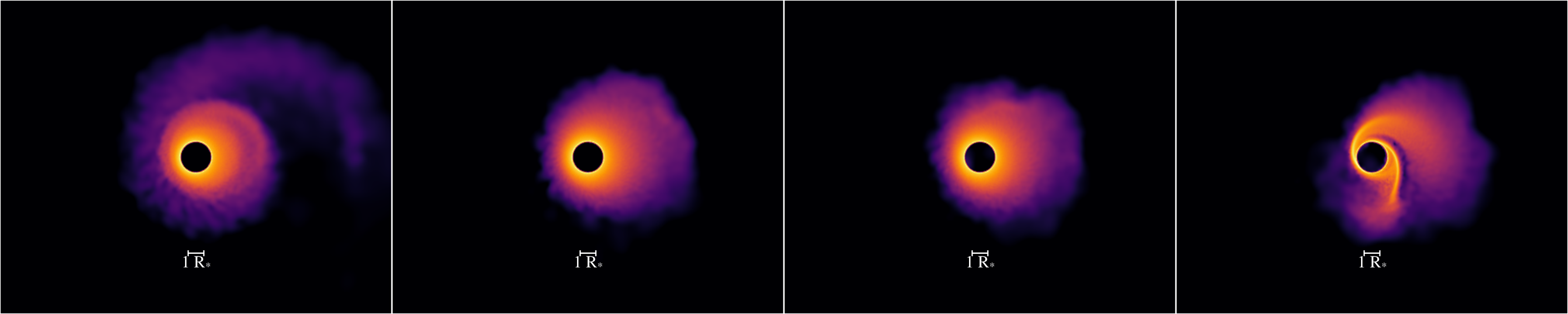}
    \caption{Misaligned by 105$^{\circ}$.}
    \end{subfigure}
\caption{Same as Fig.~\ref{fig:prograde_splash}, but for the retrograde simulations. In this view, the motion of the secondary star is clockwise. Images rendered using \textsc{SPLASH} \citep{pri07}.}
    \label{fig:retrograde_splash}
    \end{figure*}

\begin{figure*}
\centering
    \begin{subfigure}{\linewidth}
        \includegraphics[width=\linewidth]{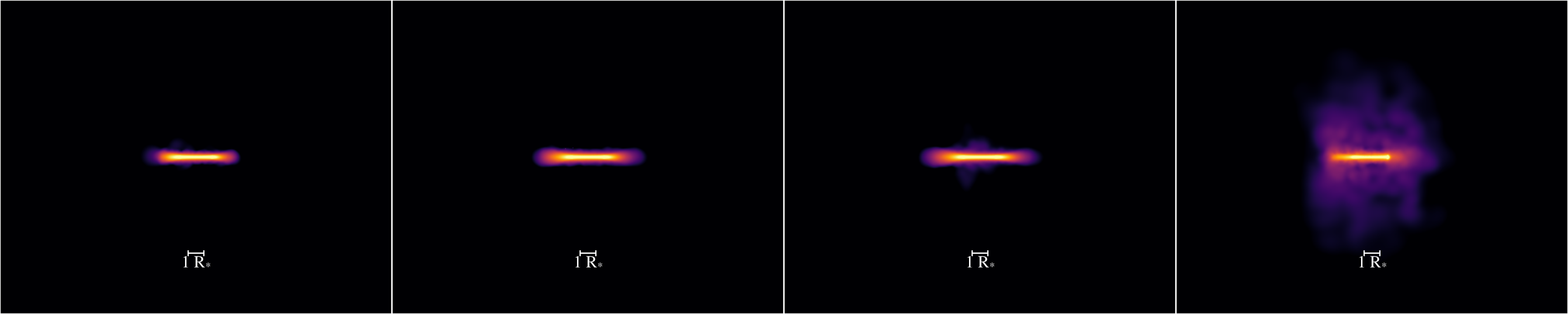}
    \caption{Misaligned by 180$^{\circ}$.}
    \end{subfigure}
\hfil
    \begin{subfigure}{\linewidth}
        \includegraphics[width=\linewidth]{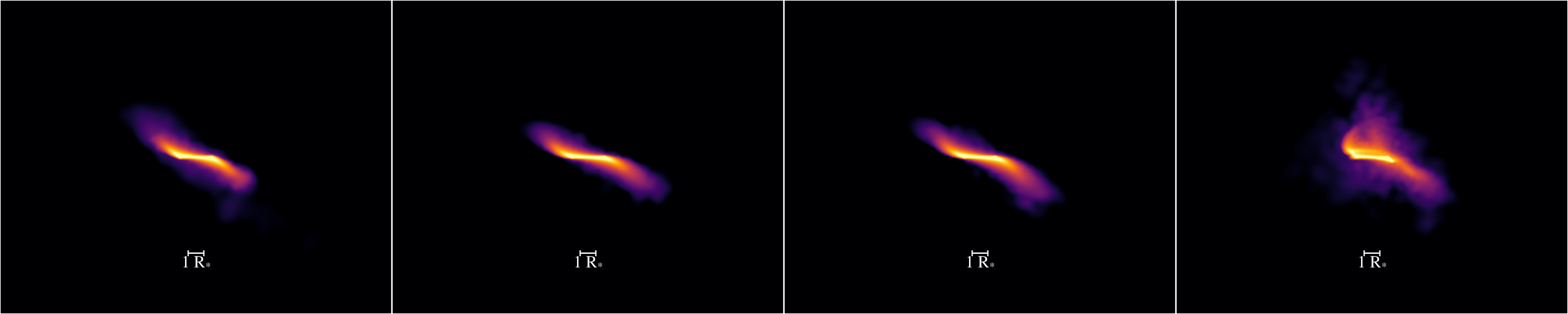}
    \caption{Misaligned by 150$^{\circ}$.}
    \end{subfigure}

    \begin{subfigure}{\linewidth}
        \includegraphics[width=\linewidth]{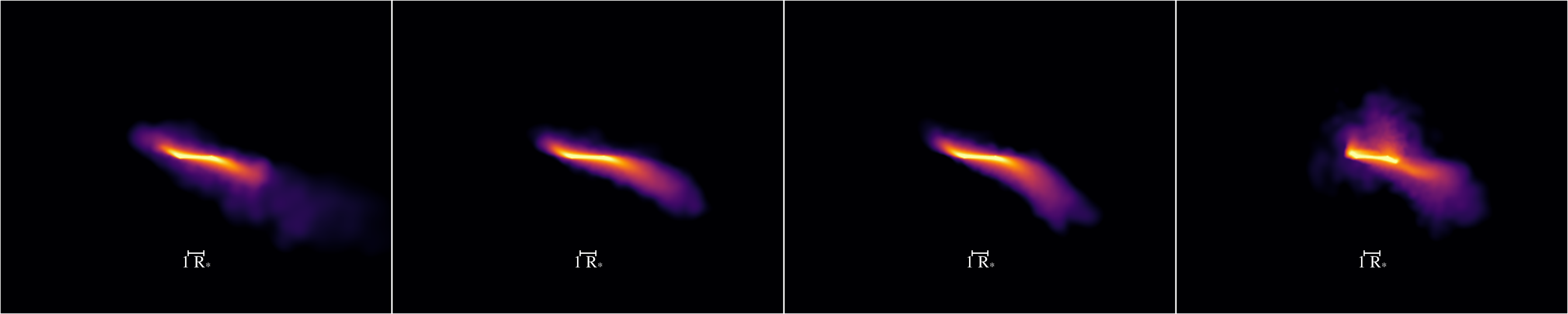}
    \caption{Misaligned by 135$^{\circ}$.}
    \end{subfigure}
\hfil
    \begin{subfigure}{\linewidth}
        \includegraphics[width=\linewidth]{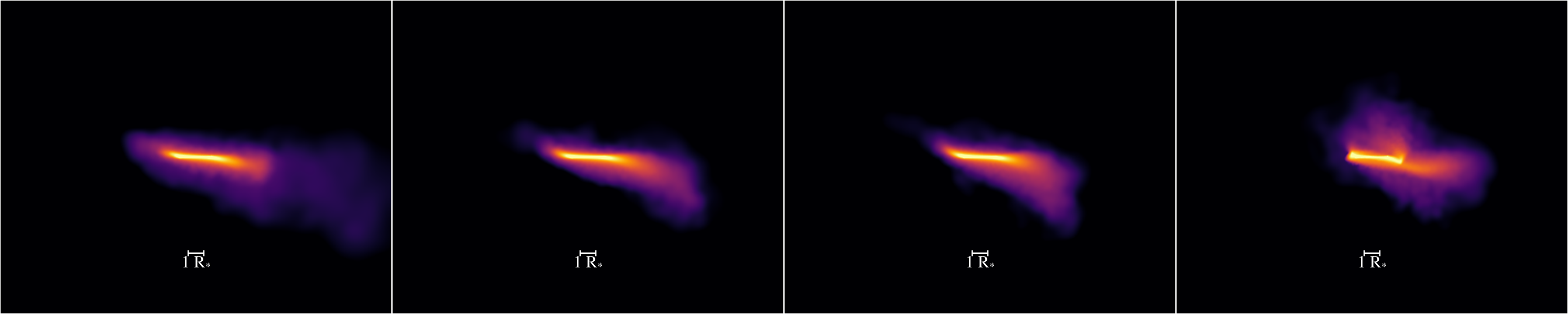}
    \caption{Misaligned by 120$^{\circ}$.}
    \end{subfigure}

    \begin{subfigure}{\linewidth}
        \includegraphics[width=\linewidth]{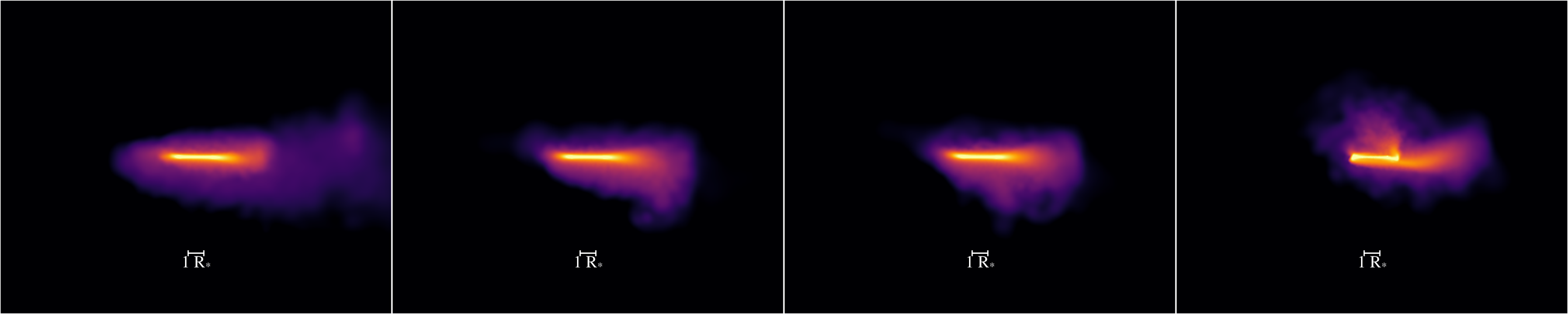}
    \caption{Misaligned by 105$^{\circ}$.}
    \end{subfigure}
\caption{Same as Fig.~\ref{fig:prograde_splash}, but for the retrograde simulations in the $x$-$z$ plane. From this perspective, the motion of the secondary star is counter-clockwise. Images rendered using \textsc{SPLASH} \citep{pri07}.}
    \label{fig:retrograde_splash_xz}
    \end{figure*}

The evolution of the retrograde discs is shown in  Fig.~\ref{fig:retro_evolution}.  Similar to the prograde models, a quasi-steady state is reached within the first ten $P_{\rm{orb}}$ and the disc's properties fluctuate with orbital phase. We see that changing the direction of the orbit from prograde to retrograde does not have a significant effect on the total disc mass in the coplanar case. However, a larger effect is seen in the misaligned systems, which as a whole have higher total disc mass. These misaligned systems do not eject as many particles into escape orbits, so more particles remain in the simulation and contribute to the total disc mass. This is further supported by the fact that the disc's average eccentricity and inclination with respect to the primary are not as responsive to the passage of the secondary in the first few encounters, indicating less disc disruption than in the prograde cases. When looking at the long-term averages for these quantities during the disc's quasi-steady state, we observe a negative correlation with misalignment angle, meaning that the disc tilts away from the primary's equatorial plane to some degree, but the magnitude of this effect is smaller than in the prograde cases.

Figure~\ref{fig:retro_coplanar_phased} shows the mass, angular momentum, eccentricity and inclination as a function of orbital phase for the coplanar retrograde simulation. We see similar trends as in the corresponding prograde simulation, with a disruption event occurring near the time of the secondary star's closest approach. We note a smaller effect on the disc's eccentricity compared to the prograde case. Figure~\ref{fig:retro_75_phased} shows the same quantities as Fig.~\ref{fig:retro_coplanar_phased}, but for the 105$^{\circ}$ case. The highly misaligned retrograde orbit shows a slightly increased disc eccentricity compared to the 180$^{\circ}$ model.   

Figure~\ref{fig:retrograde_splash} shows snapshots of the retrograde simulations in the $x$-$y$ plane at the same key moments in the orbit shown in Fig.~\ref{fig:prograde_splash}. The secondary star is unable to produce a significant accretion disc in any retrograde simulation. The effect of the neutron star on the disc is minimal for the coplanar retrograde case, with the induced asymmetries damping out completely by apastron. As we decrease the misalignment angle, the neutron star is able to induce stronger and more persistent changes to the disc. The strongest disc perturbations are seen in the 105$^{\circ}$ simulation. Fig.~\ref{fig:retrograde_splash_xz} shows snapshots of the same simulations and at the same orbital phases as Fig.~\ref{fig:retrograde_splash}, but in the $x$-$z$ plane. For misaligned simulations, we observe disc warping to become less inclined with respect to the neutron star. The only exception to this is the 105$^{\circ}$  model.

\subsubsection{Retrograde Accretion Rates}
\label{subsubsec:retrograde_accretion_rates}

\begin{figure*}
	\includegraphics[width=\textwidth]{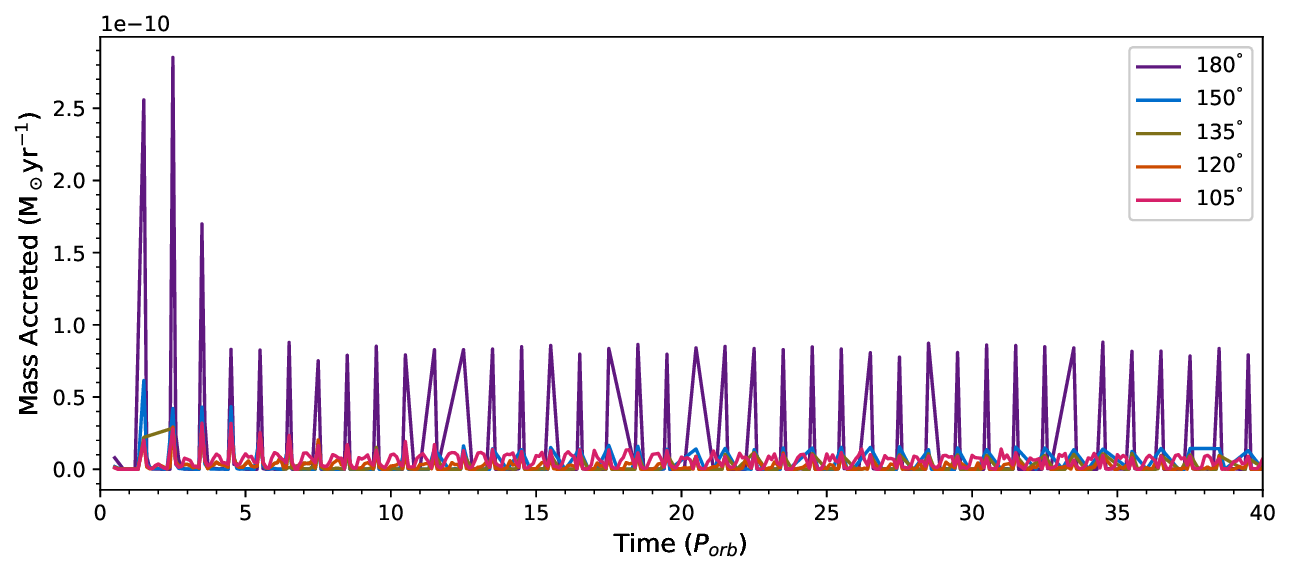}
    \caption{Same as Fig.~\ref{fig:pro_accrete_time}, but for the retrograde models.} 
    \label{fig:retro_accrete_time}
\end{figure*}

\begin{figure}
	\includegraphics[width=\columnwidth]{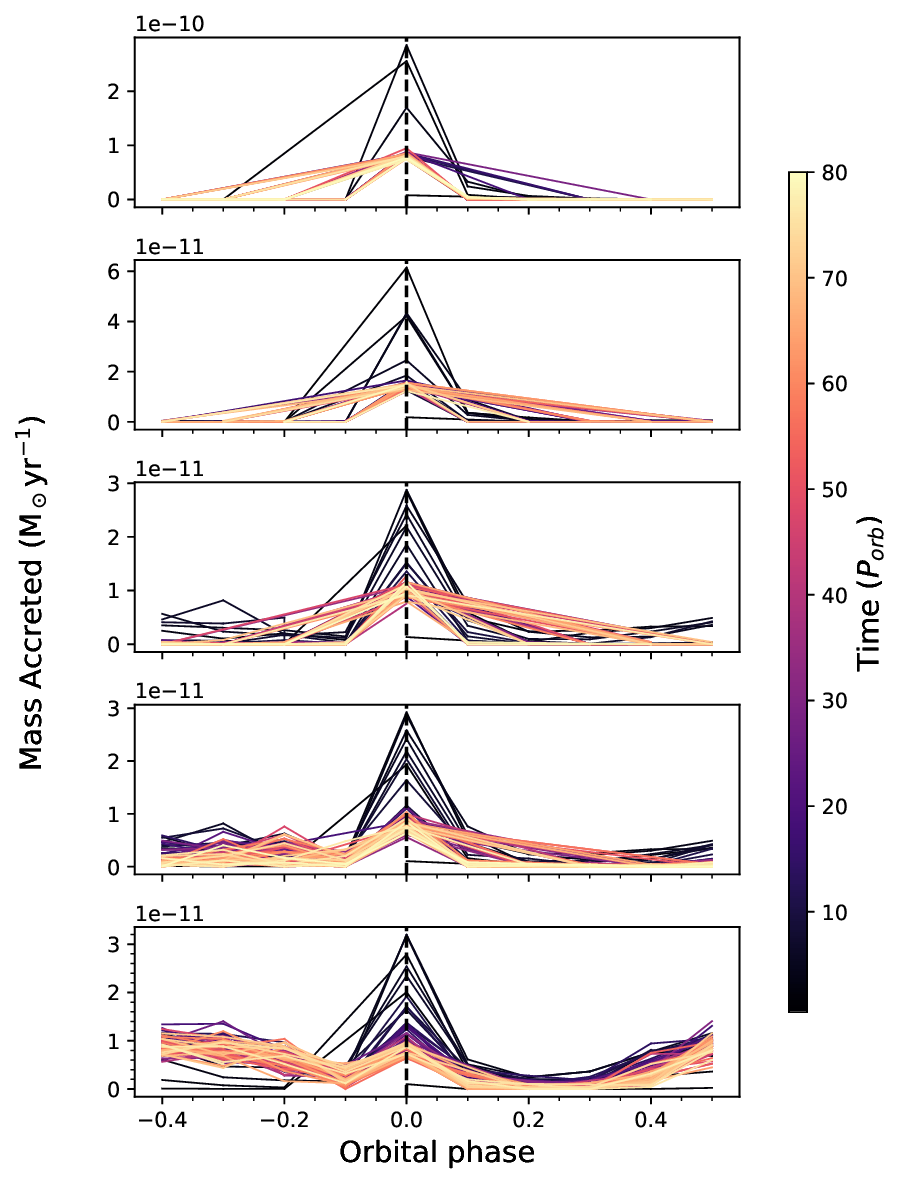}
    \caption{Same as Fig.~\ref{fig:pro_accrete_phased}, but for the retrograde models.} 
    \label{fig:retro_accrete_phased}
\end{figure}

Figure~\ref{fig:retro_accrete_time} shows the accretion rates of disc particles onto the neutron star for the retrograde models. In these simulations, the neutron star's ability to accrete matter is limited by the short interaction time between the sink particle representing the secondary, and individual disc particles. Due to their large relative velocities, far fewer disc particles remain in the Roche lobe of the companion star. In misaligned systems, the secondary star spends less time near the primary's equatorial plane, and therefore accretes even fewer particles than in the coplanar retrograde case. As a result, misalignment angles closest to coplanar retrograde are associated with the most efficient neutron star accretion.

Figure~\ref{fig:retro_accrete_phased} indicates which orbital phases are associated with the greatest accretion rates for each misalignment angle. When the secondary's orbital plane is close to the primary's equatorial plane, the accretion is most efficient at periastron (see the top three panels of Fig.~\ref{fig:retro_accrete_phased}). For misalignment angles $\leq$ 120$^{\circ}$, the neutron star is able to accrete matter at other points in the orbit, most notably during the time between apastron and periastron. 

\section{Discussion}
\label{sec:discussion}

As before, we structure our discussion to focus on the prograde models first, then the retrograde models. Our discussion ends with an evaluation of the viscosity formalism used in our simulations, including a comparison with an alternative viscosity prescription. 

\subsection{Prograde Models}
\label{subsec:prograde_discussion}

\subsubsection{Prograde Disc Evolution}
\label{subsubsec:pro_evol_discussion}

In the $x$-$y$ plane, all prograde models show the same general trends over an orbital period (see Fig.~\ref{fig:prograde_splash}). Our simulations show $m=2$ spiral density wave patterns, similar to those observed in the \textsc{SPH} simulations run by \citet{art94}, who noticed that the interference of these density waves caused significant changes in disc appearance. These spiral arms have also been observed by \citet{oka02}, \citet{pan16} and \citet{cyr17}, and explored in depth by \citet{cyr20}. The spiral arms are most defined in the 0.1 $P_{\rm{orb}}$ following periastron. Within the inner disc, the spiral density enhancements dissipate by apastron, but these features are more persistent in the outer disc. The total disc mass and angular momentum show their largest variation at the same orbital phase as the strongest spiral patterns. The coplanar case shows the largest disc disruption, losing more than 20\% of its total mass. During the short time before the next closest approach, the disc rebuilds from the inside out.

As a result of these frequent, destructive encounters with the secondary star, the disc never reaches a truly steady state. At 60.4 and 60.5 $P_{\rm{orb}}$, Fig.~\ref{fig:pro_evolution} shows that the densest, inner region on side of the disc which comes in contact with the secondary at periastron is significantly truncated. By comparison, the portion of the disc in the direction of apastron is more radially extended. This effect is especially evident for large misalignment angles. This result aligns with \citet{lub15}, who analyzed tidal torques in coplanar and misaligned circular binaries and found that misaligned discs extend further than coplanar ones, and with \citet{mir15}, who noted the same trend in eccentric systems. Similar effects have also been noted in simulations of Be/X-ray binaries produced by \citet{oka02} and \citet{oka07}, who found that the truncation effects of the binary companion on the disc are less efficient in systems with very high eccentricities.

\subsubsection{Prograde Accretion Rates}
\label{subsubsec:pro_accrete_discussion}

When considering the accretion rates on to the secondary star, two effects should be taken into account. The first is the degree to which the orbital plane of the neutron star overlaps with that of the primary's equatorial plane. This controls how much of the disc the neutron star passes through and therefore the number of particles the neutron star encounters. The second is the relative velocities of the disc particles and the neutron star. In the prograde models, these effects work together as additive effects. For example, we see that accretion rates are negatively correlated with misalignment angle in Fig.~\ref{fig:pro_accrete_time}. In the coplanar case, the neutron star passes through disc for longer, interacting with a larger number of particles, and is able to gravitationally capture more. The relative velocities of the neutron star and the disc particles are also small, since the velocity vectors are pointed in roughly the same direction. Therefore, the neutron star has relatively long interaction times with each particle, resulting in a high accretion rate. Increasing the misalignment angle results in decreasing the total number of particles that the secondary star interacts with, and also reduces the component of the neutron star's velocity vector that is parallel to the disc particles'. This results in lower overall accretion rates with increasing misalignment angle.

The accretion rates over an orbital period, shown in Fig.~\ref{fig:pro_accrete_phased}, can be explained in terms of system geometry and the disc asymmetries created by the neutron star. For the coplanar prograde model, we see peak accretion rates at periastron and shortly after. As we see in the top panel of Fig.~\ref{fig:prograde_splash}, the neutron star passes through the densest, innermost regions of the disc at this orbital phase, and is therefore is able to efficiently accumulate matter at rates that are roughly twice as large as the other tested misaligned cases. The neutron star induces a spiral arm as it moves away from the primary after periastron. The contact with this spiral arm enables the neutron star to maintain its peak accretion rate until about 0.1 $P_{\rm{orb}}$ after periastron. The models with moderate misalignment angles, shown in the second and third panels of Fig.~\ref{fig:pro_accrete_phased}, are able to accrete material most efficiently after periastron once the discs have matured. For these systems, the peak accretion rates occur when the spiral arm structure is at its strongest. Panels (b) and (c) of Fig.~\ref{fig:prograde_splash_xz} show that the disc has warped and produced a strong accretion stream connecting the innermost disc to the neutron star. 

For the models with highly misaligned binary orbits, we might expect two peak accretion times to occur; one before periastron and the other after, since the orbital path of the neutron star intersects the plane of the primary's disc twice at these points. However, the 60$^{\circ}$ and 75$^{\circ}$ models in Fig.~\ref{fig:pro_accrete_phased} show two peaks, one coinciding with periastron, and the second occurring within 0.2 or 0.3 $P_{\rm{orb}}$ afterward. These observations can be explained intuitively by a comparison of the bottom four panels in Figs.~\ref{fig:prograde_splash} and~\ref{fig:prograde_splash_xz}. At the moment 0.1 $P_{\rm{orb}}$ before periastron, the neutron star has not yet entered the primary's disc. At periastron, the secondary star is again able to interact with the dense regions of the inner disc. For the 60$^{\circ}$ and 75$^{\circ}$ models, the companion star quickly gains vertical distance from the plane of the disc and therefore the accretion streams shown in panels (d) and (e) of Fig.~\ref{fig:prograde_splash_xz} are not as strong as those seen in the 30$^{\circ}$ and 45$^{\circ}$ models. This effect is especially pronounced for the 75$^{\circ}$ models. However, the arm of disc material connecting the secondary to the primary's disc survives until periastron, as seen in panels (d) and (e) of Fig.~\ref{fig:prograde_splash}. This arm of material accounts for the larger accretion rate between apastron and periastron, which is clear in the bottom two panels of Fig.~\ref{fig:pro_accrete_phased}.

\subsection{Retrograde Models}
\label{subsubsec:retro_evol_discussion}

When comparing the effects of larger misalignment angles for retrograde models, we see the opposite trend compared to prograde models. In Fig.~\ref{fig:retrograde_splash}, we see that decreasing the misalignment angle results in strong disc disruption. This can be explained by the velocity vectors of the neutron star and the disc particles. For the coplanar retrograde model, the neutron star's velocity vector is directed completely antiparallel to the disc's. As a result, the relative velocities between the neutron star and the individual disc particles are very large. As we decrease the misalignment angle of the neutron star's orbit, we increase a component of the velocity vector that is perpendicular, rather than antiparallel, to the disc particles' velocities. As a result, the relative velocities are smaller and the neutron star can interact with individual particles for longer. This produces larger disc disruptions, more eccentric discs, as well as two-armed spiral density patterns for the smallest retrograde misalignment angles.  

\subsubsection{Retrograde Accretion Rates}
\label{subsubsec:retro_accrete_discussion}

When considering the effects of misalignment angle on accretion rates shown in Fig.~\ref{fig:retro_accrete_time}, the two effects mentioned in Section~\ref{subsubsec:pro_accrete_discussion} are still occurring, but rather than working together, they are competing. In the coplanar retrograde case, the neutron star's path overlaps with the paths of the particles in the disc much more than it does for the misaligned cases. As a result, a larger number of disc particles are able to interact with the neutron star and enter its Roche lobe. Therefore, even though the relative velocities between the neutron star and the disc particles are at their largest in the coplanar retrograde case, the total number of particles that can enter the Roche lobe is larger, and therefore accretion rates are higher than in the other retrograde cases. For smaller misalignment angles, the neutron star interacts with fewer particles as it intersects the disc at specific points, rather than travelling through its plane. So, the total number of particles in the proximity of the secondary limits the number that can be accreted. 

In order to explain the accretion over an orbital period, it is again useful to consider the asymmetries of the disc generated by the neutron star. In the coplanar case, the peak accretion rates occur at periastron, as seen in the first panel of Fig.~\ref{fig:retro_accrete_phased}. The secondary does not induce spiral arms into the disc in this case, nor does it move with a significant portion of the disc as it travels away from its closest approach. Therefore, the neutron star is unable to efficiently capture particles at any other point in the orbital period. The misaligned cases also show strong peak accretion rates at periastron early in the system's evolution, but this peak becomes less pronounced over time. As the misalignment angle decreases, more matter is stripped from the disc by the passage of the secondary star and stronger spiral density patterns are induced. As we see in panels (c), (d) and (e) of Fig.~\ref{fig:retrograde_splash}, this matter is flung in the direction of apastron. Panels (c), (d) and (e) of Fig.~\ref{fig:retrograde_splash_xz}, also show that the disc is warping. The path of the secondary star intersects these arms of material between apastron and periastron. Therefore, during this portion of its orbit, the neutron star is able to interact with a larger number of particles than in cases without this spiral arm structure. Some of these particles enter the Roche lobe and are accreted, leading to the increased accretion rates between apastron and periastron seen in the bottom three panels of Fig.~\ref{fig:retro_accrete_phased}.

\subsection{Comparing viscosity prescriptions}

\begin{figure}
\centering
    \begin{subfigure}{\linewidth}
        \includegraphics[width=\linewidth]{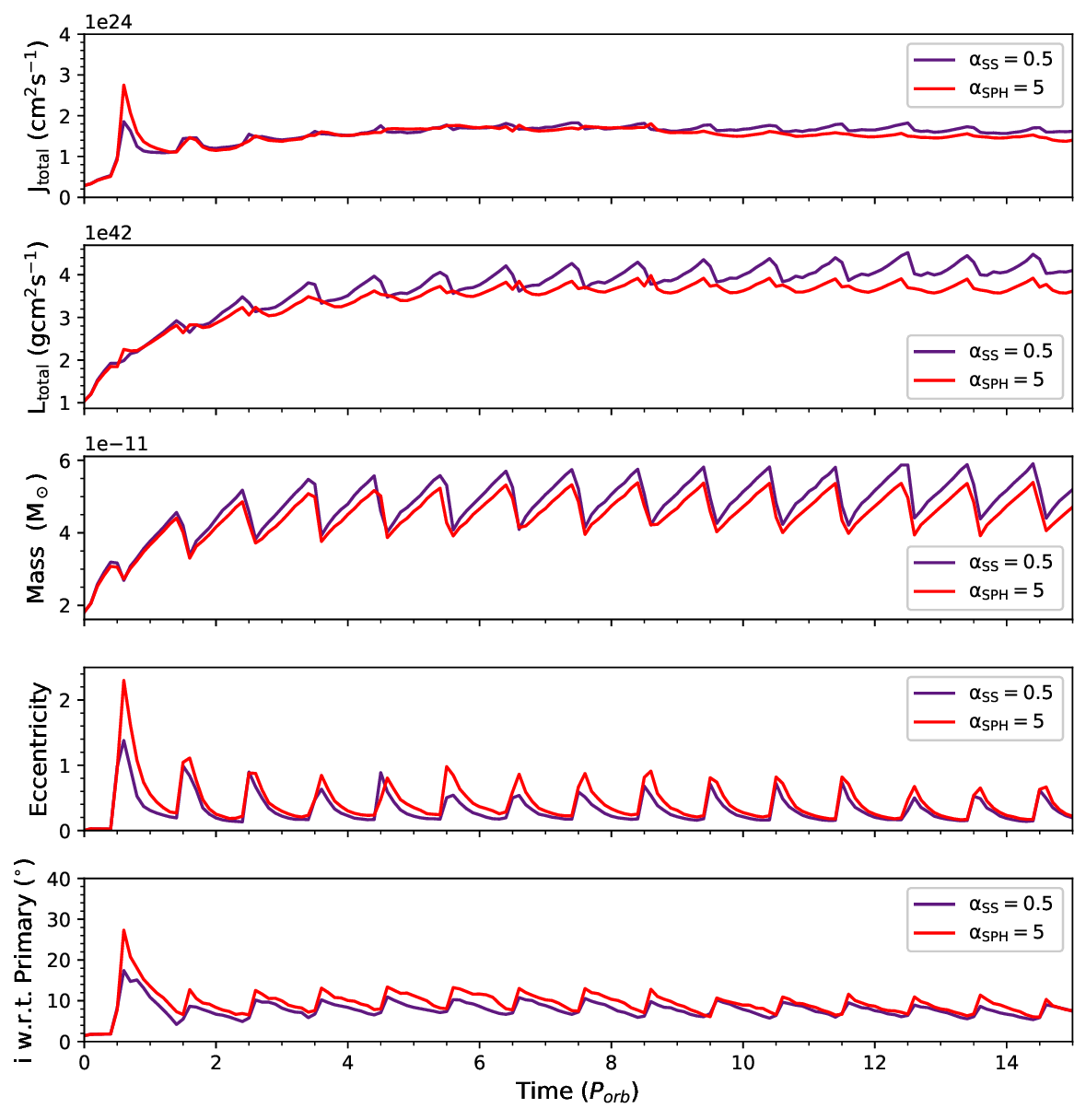}
    \caption{Misaligned by 0$^{\circ}$.}
    \end{subfigure}
\hfil
    \begin{subfigure}{\linewidth}
        \includegraphics[width=\linewidth]{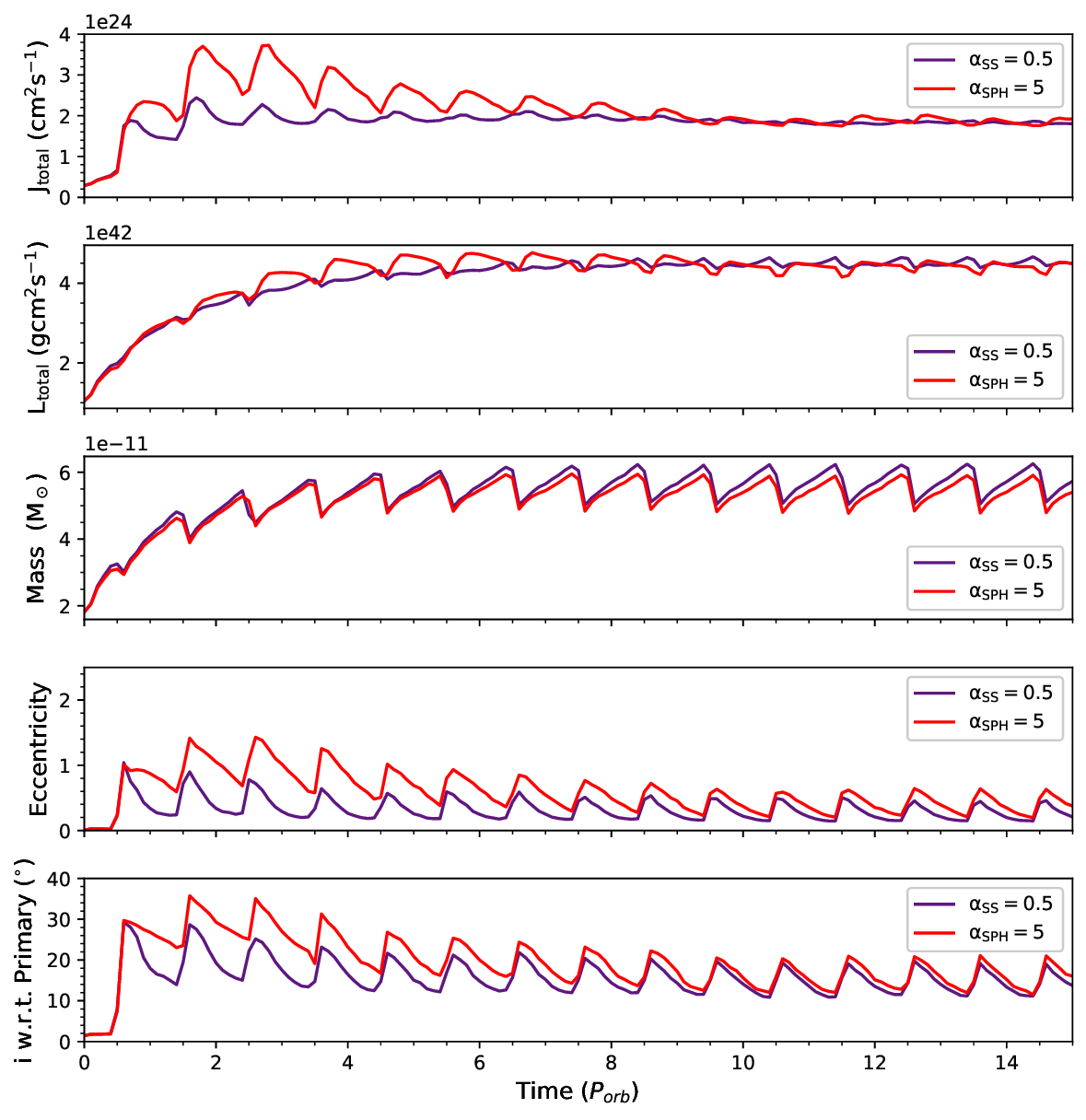}
    \caption{Misaligned by 45$^{\circ}$.}
    \end{subfigure}
\caption{Comparison of the disc evolution using $\alpha_{\rm{SS}}$ and $\alpha_{\rm{SPH}}$ for the prograde models. Panel (a) represents the typical observed difference, while panel (b) represents the largest difference.}
    \label{fig:alpha_pro_evol_comp}
    \end{figure}

\begin{figure}
	\includegraphics[width=\columnwidth]{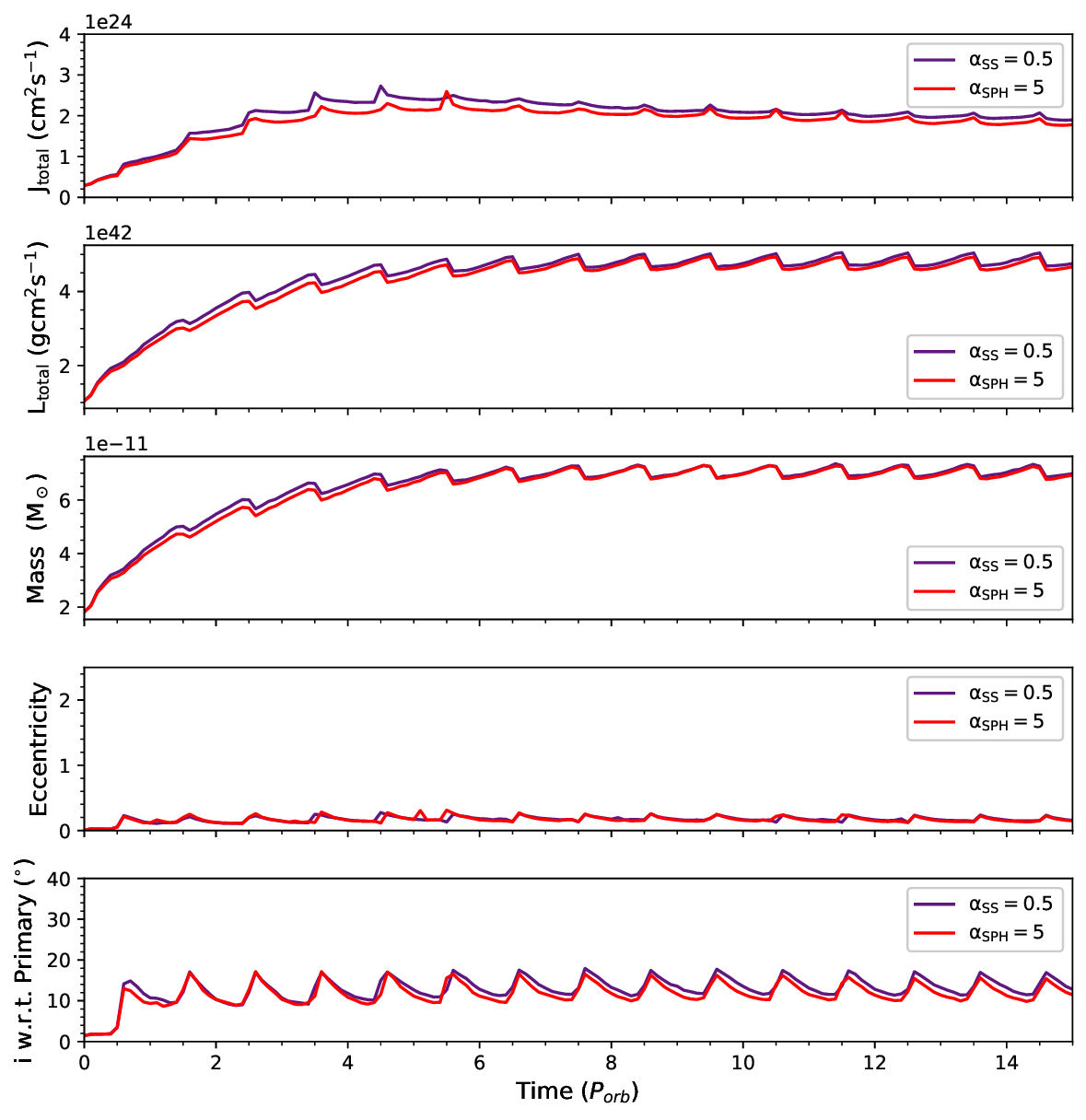}
    \caption{Comparison of the disc evolution using $\alpha_{\rm{SS}}$ and $\alpha_{\rm{SPH}}$ for the 135$^{\circ}$ model. This behaviour is representative of all tested retrograde models.} 
    \label{fig:alpha_retro_evol_comp}
\end{figure}

\begin{figure}
\centering
    \begin{subfigure}{\linewidth}
        \includegraphics[width=\linewidth]{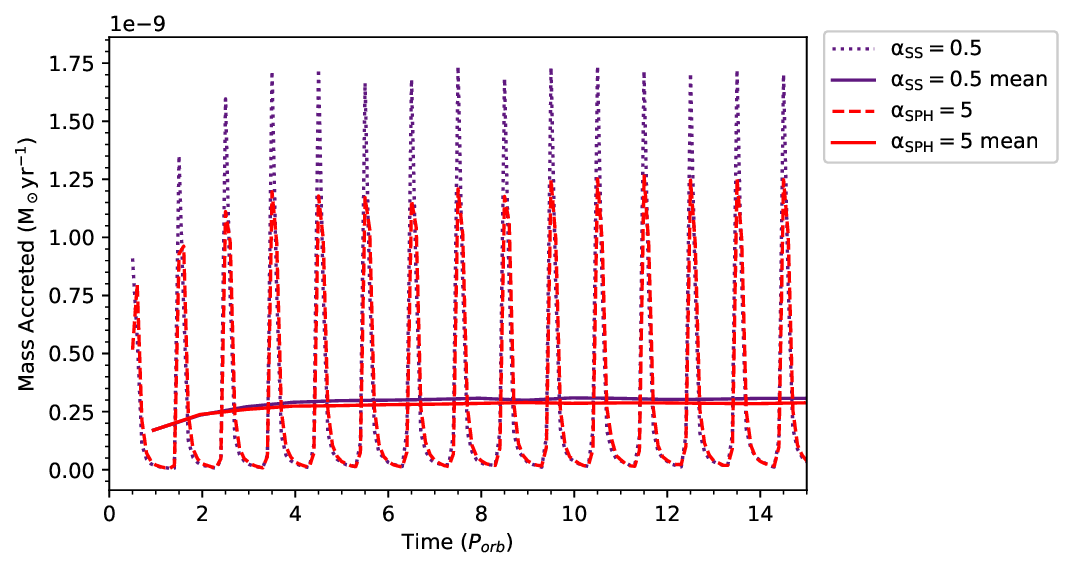}
    \caption{Misaligned by 0$^{\circ}$.}
    \end{subfigure}
\hfil
    \begin{subfigure}{\linewidth}
        \includegraphics[width=\linewidth]{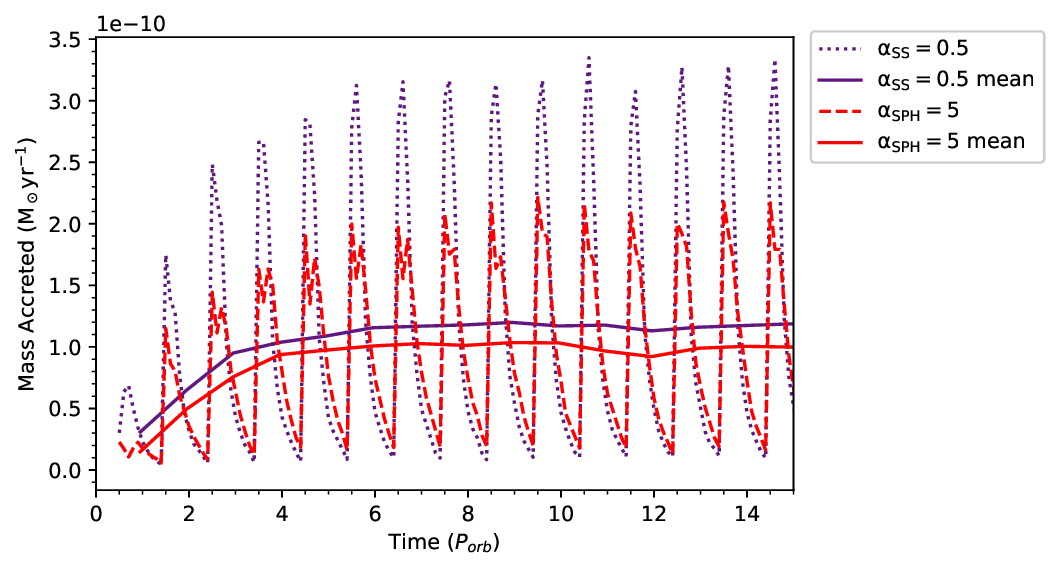}
    \caption{Misaligned by 45$^{\circ}$.}
    \end{subfigure}
\caption{Comparison of the accretion rates using $\alpha_{\rm{SS}}$ and $\alpha_{\rm{SPH}}$ for the prograde models. Panel (a) represents the typical observed difference, while panel (b) represents the largest difference. The solid line indicates the mean accretion rate over 1 $P_{\rm{orb}}$.}
    \label{fig:alpha_pro_comp}
    \end{figure}

\begin{figure}
\centering
    \begin{subfigure}{\linewidth}
        \includegraphics[width=\linewidth]{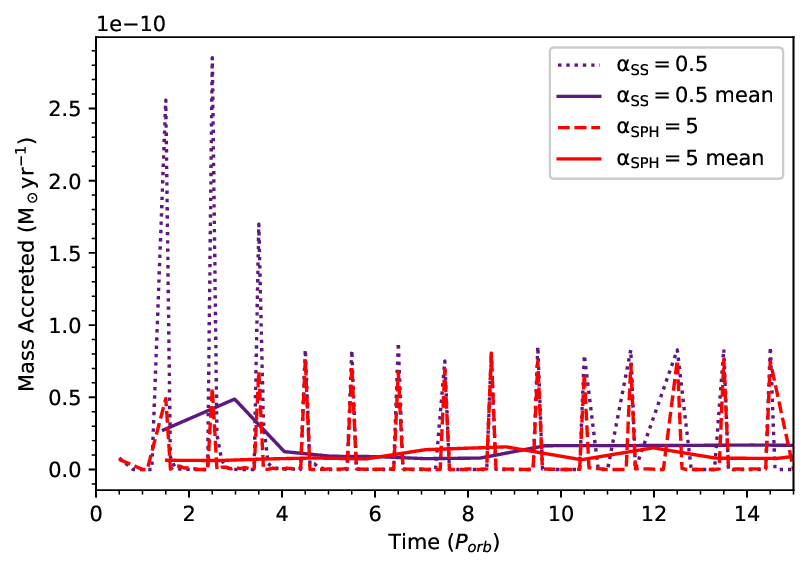}
    \caption{Misaligned by 180$^{\circ}$.}
    \end{subfigure}
\hfil
    \begin{subfigure}{\linewidth}
        \includegraphics[width=\linewidth]{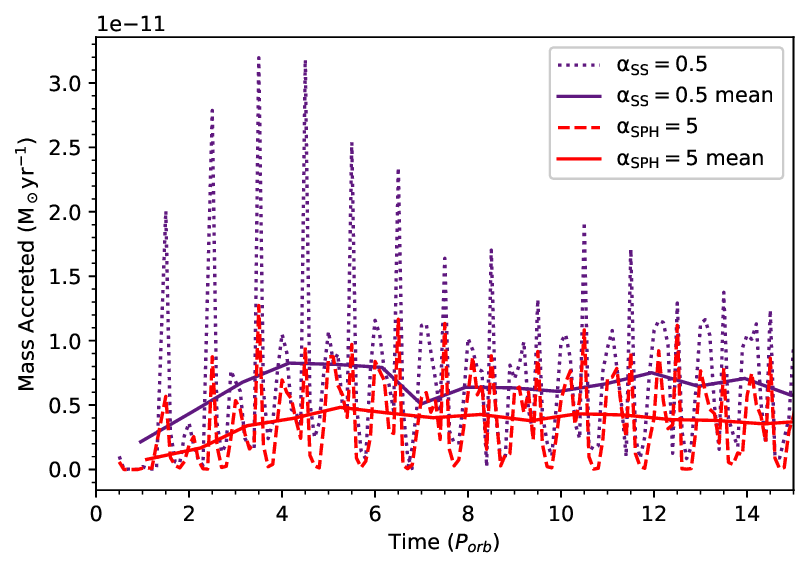}
    \caption{Misaligned by 105$^{\circ}$.}
    \end{subfigure}
\caption{Same as Fig.~\ref{fig:alpha_pro_comp}, but for the retrograde models.}
    \label{fig:alpha_retro_comp}
    \end{figure}

As discussed in Section~\ref{sec:methods}, our models utilise the Shakura-Sunyaev prescription for the viscosity which depends on the dimensionless parameter $\alpha_{\rm{SS}}$. This parameter is held constant in space and time. As we see in Equation~\ref{eq:sph_viscosity}, $\alpha_{\rm{SS}}$ is related to the artificial viscosity parameter used internally by our \textsc{SPH} code ($\alpha_{\rm{SPH}}$) through a scaling factor and the ratio between the smoothing length and the scale height. The scale height is calculated assuming an isothermal disc and uses the radial distance $r$ of a particle from the primary star 
\begin{equation}
    H(r) = \frac{c_s R_{\star}}{v_{\rm{crit}}} \left( \frac{r}{R_{\star}} \right)^{1.5} ,\\
    \label{eq:scale_height}
\end{equation}
where $c_s$ is the sound speed in the disc and $v_{\rm{crit}}=\left(G M/R_{\star}\right)^{1/2}$ is the critical velocity of the star \citep{car06a}. As the ratio between the smoothing length and the scale height changes throughout the disc, $\alpha_{\rm{SPH}}$ must adapt in order to hold $\alpha_{\rm{SS}}$ constant. For particles that are gravitationally bound to the primary star, this approximation of the scale height holds and the application of Equation~\ref{eq:scale_height} is justified. However, for particles bound to the secondary star, the scale height does not truly follow the relation in Equation~\ref{eq:scale_height} and is overestimated when this equation is applied. In response, $\alpha_{\rm{SPH}}$ must also increase in order to maintain a constant $\alpha_{\rm{SS}}$ in those regions. This affects the interactions between the particles near the secondary star and therefore their accretion timescales. This effect may be especially significant for misaligned systems. 

To explore the extent to which our findings are affected by our choice of viscosity prescription, we tested a subset of our models with a variable viscosity parameter. Rather than defining our viscosity through $\alpha_{\rm{SS}}$ which must be held constant in space and time, we chose a value of $\alpha_{\rm{SPH}}$ which is roughly equivalent. Following the relationship in Equation~\ref{eq:sph_viscosity}, we set $\alpha_{\rm{SPH}}$ = 5, which should approximate the $\alpha_{\rm{SS}}$ = 0.5 used in our other models. We tested 0$^{\circ}$, 45$^{\circ}$, 75$^{\circ}$, 105$^{\circ}$, 135$^{\circ}$, and 180$^{\circ}$ models.

We find a good correspondence between the disc evolution in both viscosity regimes. In the 0$^{\circ}$ and 75$^{\circ}$ models, the discs evolve almost identically with $\alpha_{\rm{SPH}}$ = 5 compared to $\alpha_{\rm{SS}}$ = 0.5. In the 45$^{\circ}$ model, larger variations are noticed in the first five orbital periods, but these become less significant as the disc matures. A comparison between the typical observed comparison, and the largest differences we find, are shown in Fig.~\ref{fig:alpha_pro_evol_comp} for the prograde models. Among the retrograde models, both viscosity prescriptions produce the same results in all tested simulations, so we do not provide a comparison between the typical and largest observed differences. Instead, Fig.~\ref{fig:alpha_retro_evol_comp} is representative of all three tested retrograde models. Overall, our choice of a viscosity parameter does not alter the observed trends in large-scale disc evolution.   

Next, we consider the effect of a variable viscosity parameter on accretion rates. We find that using the artificial viscosity prescription with $\alpha_{\rm{SPH}}$ = 5 results in lower accretion rates across all models. The extent of the difference is dependent on the misalignment angle in both prograde and retrograde cases. The largest observed effects on accretion rates are seen in the 45$^{\circ}$ and 105$^{\circ}$ models. However, the differences are below 30\% for most models. Figure.~\ref{fig:alpha_pro_comp} shows a typical comparison of accretion rates for both viscosity prescriptions in panel (a), and the model with the largest observed difference in panel (b). Figure~\ref{fig:alpha_retro_comp} shows the same values, but for the retrograde models. We see that changing the viscosity prescription does not change the order-of-magnitude estimate of accretion rates, and both scenarios show the same approximate trends over the course of an orbital period.

\subsubsection{Effects of varying the longitude of the ascending node}
\label{subsubsec:var_omega}

All models reported in this work defined the same longitude of the ascending node for the binary orbit ($\Omega = 0^{\circ}$). We expect the effects of varying this parameter to be strongest for models with large misalignment angles. To test the effects of this parameter, we compared the 75$^{\circ}$ model with $\Omega = 0^{\circ}$ to $\Omega = 90^{\circ}$. Over ten orbital periods, the mass and total angular momentum of the disc in the model with $\Omega = 90^{\circ}$ were no more than two times larger than in the model with $\Omega = 0^{\circ}$. The eccentricity and inclinations with respect to the primary and secondary stars were not significantly affected. In addition, no impact was observed on the measured trends with orbital phase.

\section{Conclusions}
\label{sec:conclusion}

We investigated the large-scale Be star disc evolution in highly eccentric Be/X-ray binaries in a variety of system geometries. Our models are motivated by A0538-66, a short-period Be/X-ray binary with an eccentricity of 0.72. We fix all parameters at values consistent with this system, except for the misalignment angle between the spin axis of the primary star and the neutron star's orbital plane, which we vary from a coplanar prograde case of 0$^{\circ}$ to the coplanar retrograde case of 180$^{\circ}$}. We vary this parameter to create ten models which explore a wide range of misaligned models, in steps.

We find that the secondary star induces $m=2$ spiral density enhancements in the Be star disc among almost all tested geometries. In all models, the high eccentricity of the binary's orbit causes the calculated properties of Be star disc to vary with orbital phase. For prograde models, larger misalignment angles result in smaller variations in disc mass, angular momentum, eccentricity, and inclinations relative to the primary and secondary's spin axes. For these models, larger misalignment angles are also associated with less defined spiral arms and lower accretion rates onto the neutron star. This effect is due to the magnitude of the component of the neutron star's velocity vector that is \textit{parallel} to that of the disc particles. The accretion rates can be explained by considering the number of particles the neutron star can interact with, as well as the length of interaction time which is limited by the relative velocities of the neutron star and individual disc particles. Accretion rates also vary with orbital phase, reaching peak levels at different points in the orbit depending on the level of disc disruption and the number of particles near the secondary star at a given orbital phase. 

Among the retrograde models, smaller misalignment angles result in stronger disc disruptions and the formation of more distinct spiral structures. This effect is due to the magnitude of the component of the neutron star's velocity vector that is \textit{antiparallel} to that of the disc particles. For the coplanar retrograde case, the velocity vector of the neutron star has a large component directed antiparallel to the motion of the disc particles, limiting the interaction time with each particle. Smaller misalignment angles reduce the magnitude of this component of the velocity vector and therefore increase the reaction times between the neutron star and individual particles, inducing more significant disruptive effects in the disc. As with the prograde models, the accretion rates in the retrograde models can be explained in terms of the number of particles interacting with the neutron star, and the length of the interaction time. Over an orbital period, retrograde models with close to coplanar show maximum accretion rates at periastron, while those with large misalignment angles can interact with spiral structures in the disc at other orbital phases, allowing them to accrete over a larger portion of the orbit. 

This work provides an analysis of the physical characteristics of the decretion discs in Be/X-ray systems with extreme orbits. Our estimated rates of accretion by the neutron star can be used to calculate X-ray flux values expected for each orbital configuration. We will explore this in a future paper, along with the predicted observational signatures of the Be decretion and neutron star accretion discs. 

\section*{Acknowledgements}

C.E.J. acknowledges support through the National Science and Engineering Research Council of Canada. A.C.C. acknowledges support from CNPq (grant 314545/2023-9) and FAPESP (grants 2018/04055-8 and 2019/13354-1). This work was made possible through the use of the Shared Hierarchical Academic Research Computing Network (SHARCNET). We acknowledge the use of SPLASH \citep{pri07} for rendering and visualisation of our figures. We acknowledge the Anishinaabek, Haudenosaunee,  L\={u}naap\'ewak and Chonnonton Nations, on whose traditional territories this work was produced.

\section*{Data Availability}

No new data were generated or analysed in support of this research.



\bibliographystyle{mnras}
\bibliography{example} 




\appendix

\section{Phased Disc Evolution and Bound/Unbound Particles}
\label{sec:appendix}

Here, we provide supplementary figures for the large-scale disc evolution as a function of orbital phase, for the simulations not shown in the main text. These include the models with misalignment angles of 30$^{\circ}$, 45$^{\circ}$, 60$^{\circ}$, 120$^{\circ}$, 135$^{\circ}$, and 150$^{\circ}$ models.   

\begin{figure}
	\includegraphics[width=\columnwidth]{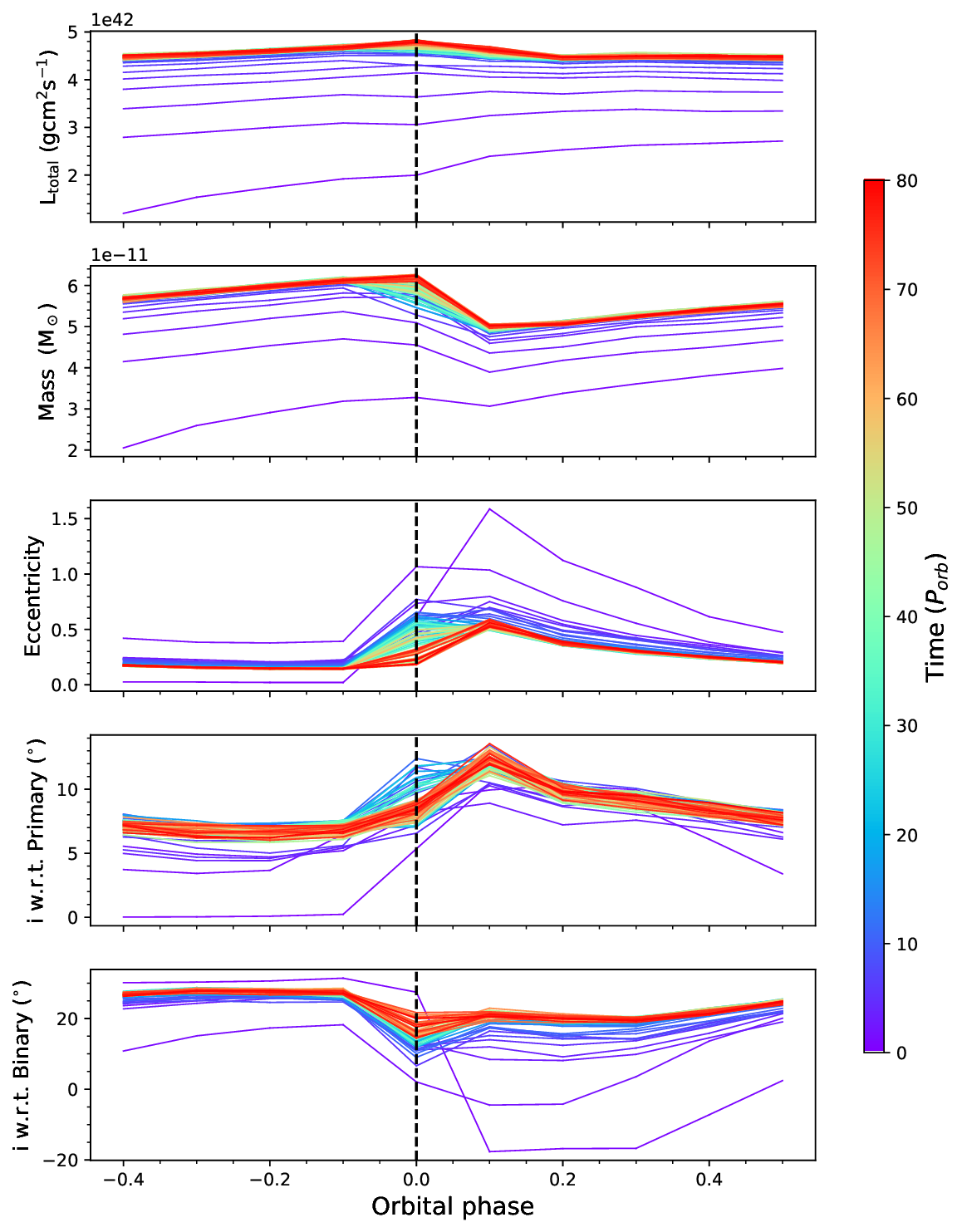}
    \caption{Same as Fig.~\ref{fig:pro_coplanar_phased}, but for the 30$^{\circ}$ simulation.} 
    \label{fig:pro_30_phased}
\end{figure}

\begin{figure}
	\includegraphics[width=\columnwidth]{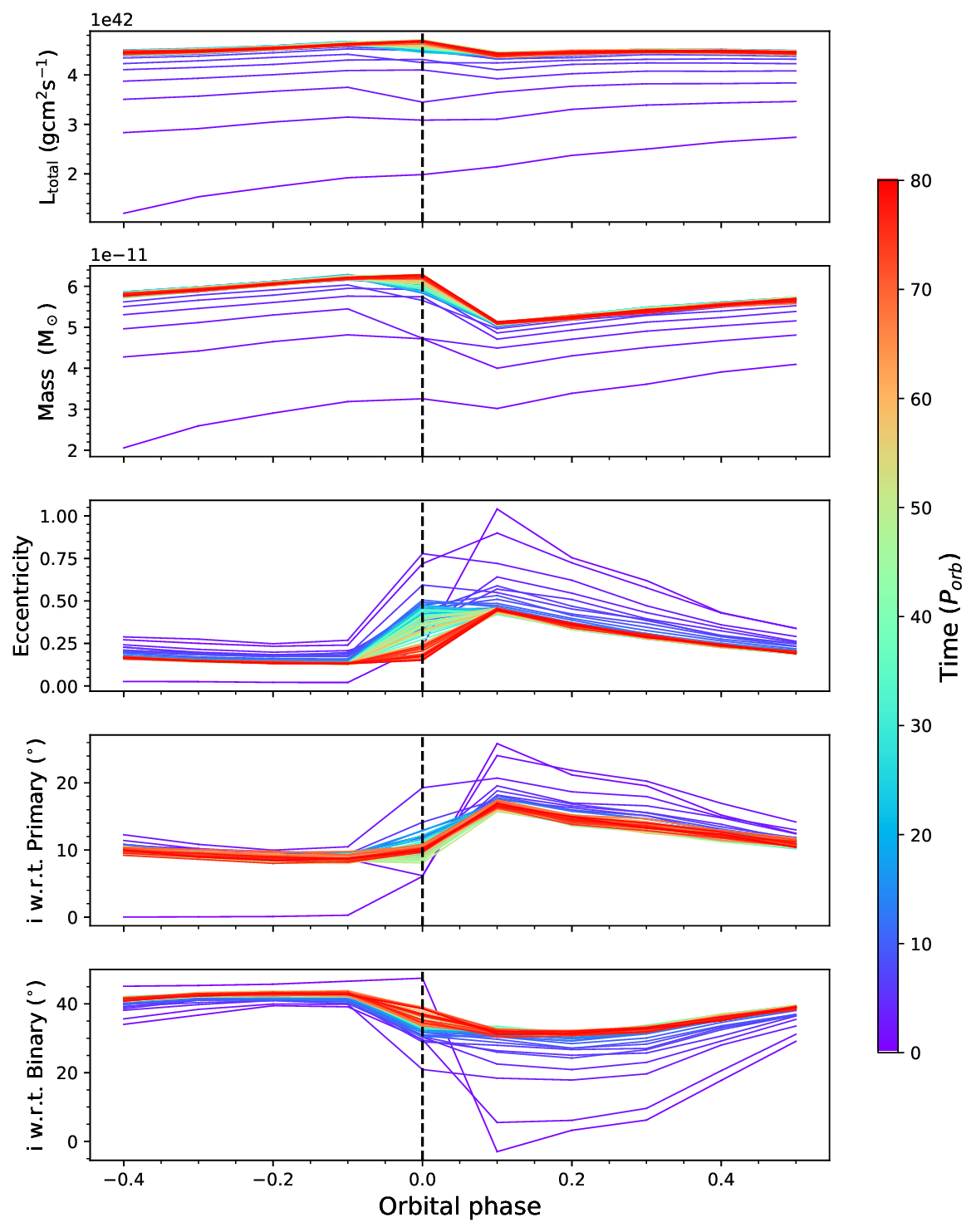}
    \caption{Same as Fig.~\ref{fig:pro_coplanar_phased}, but for the 45$^{\circ}$ simulation.} 
    \label{fig:pro_45_phased}
\end{figure}

\begin{figure}
	\includegraphics[width=\columnwidth]{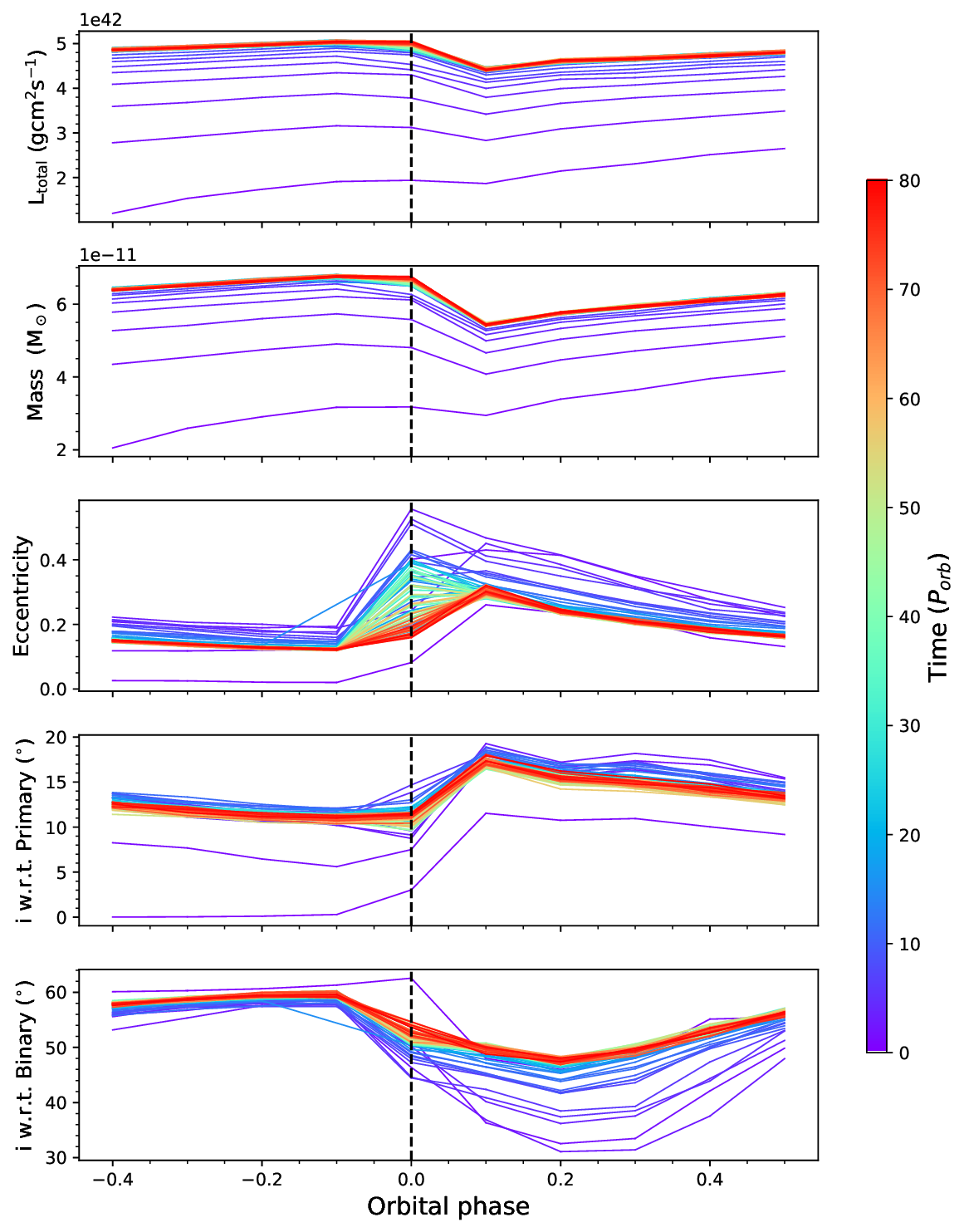}
    \caption{Same as Fig.~\ref{fig:pro_coplanar_phased}, but for the 60$^{\circ}$ simulation.} 
    \label{fig:pro_60_phased}
\end{figure}

\begin{figure}
	\includegraphics[width=\columnwidth]{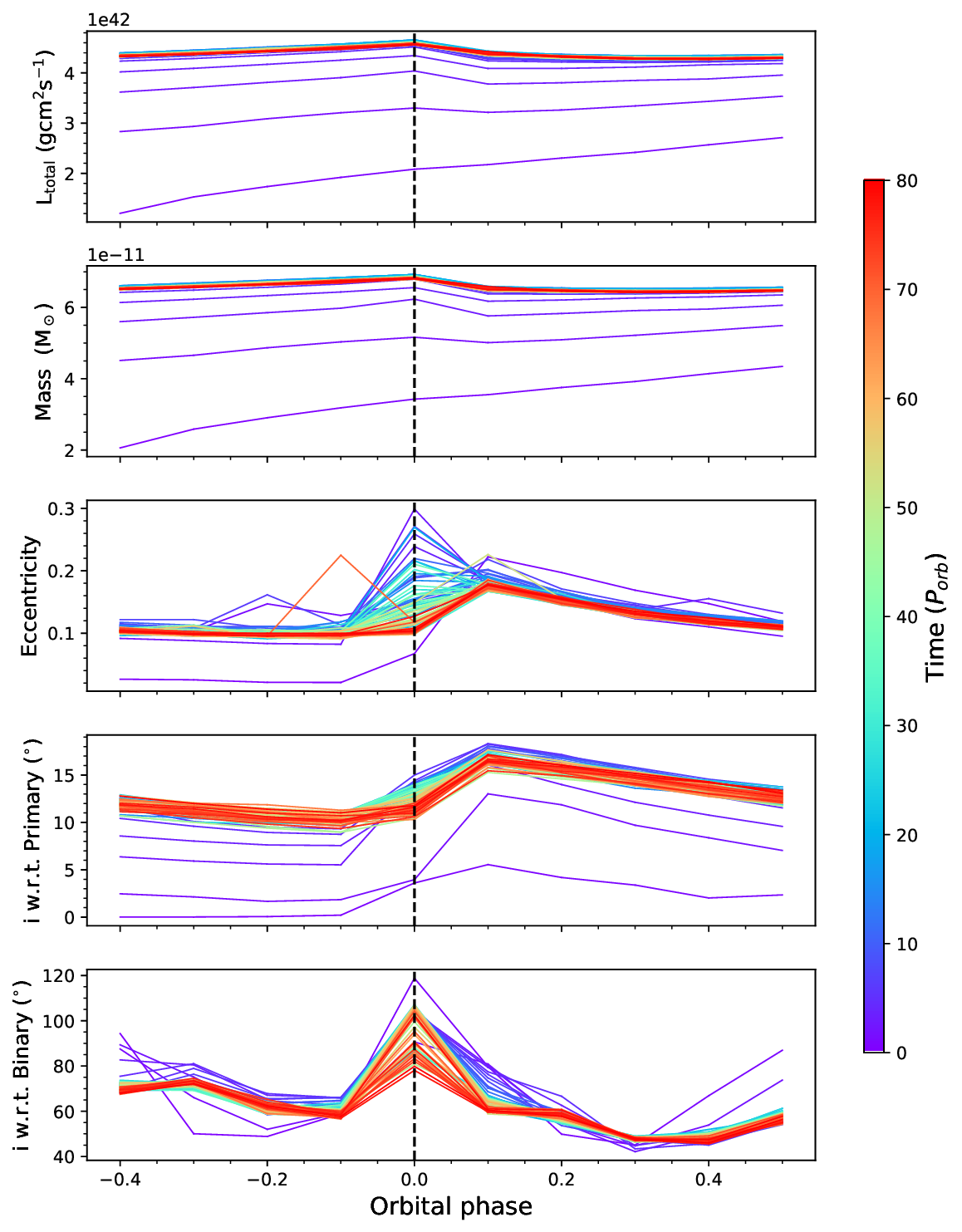}
    \caption{Same as Fig.~\ref{fig:pro_coplanar_phased}, but for the 150$^{\circ}$ simulation.} 
    \label{fig:retro_30_phased}
\end{figure}

\begin{figure}
	\includegraphics[width=\columnwidth]{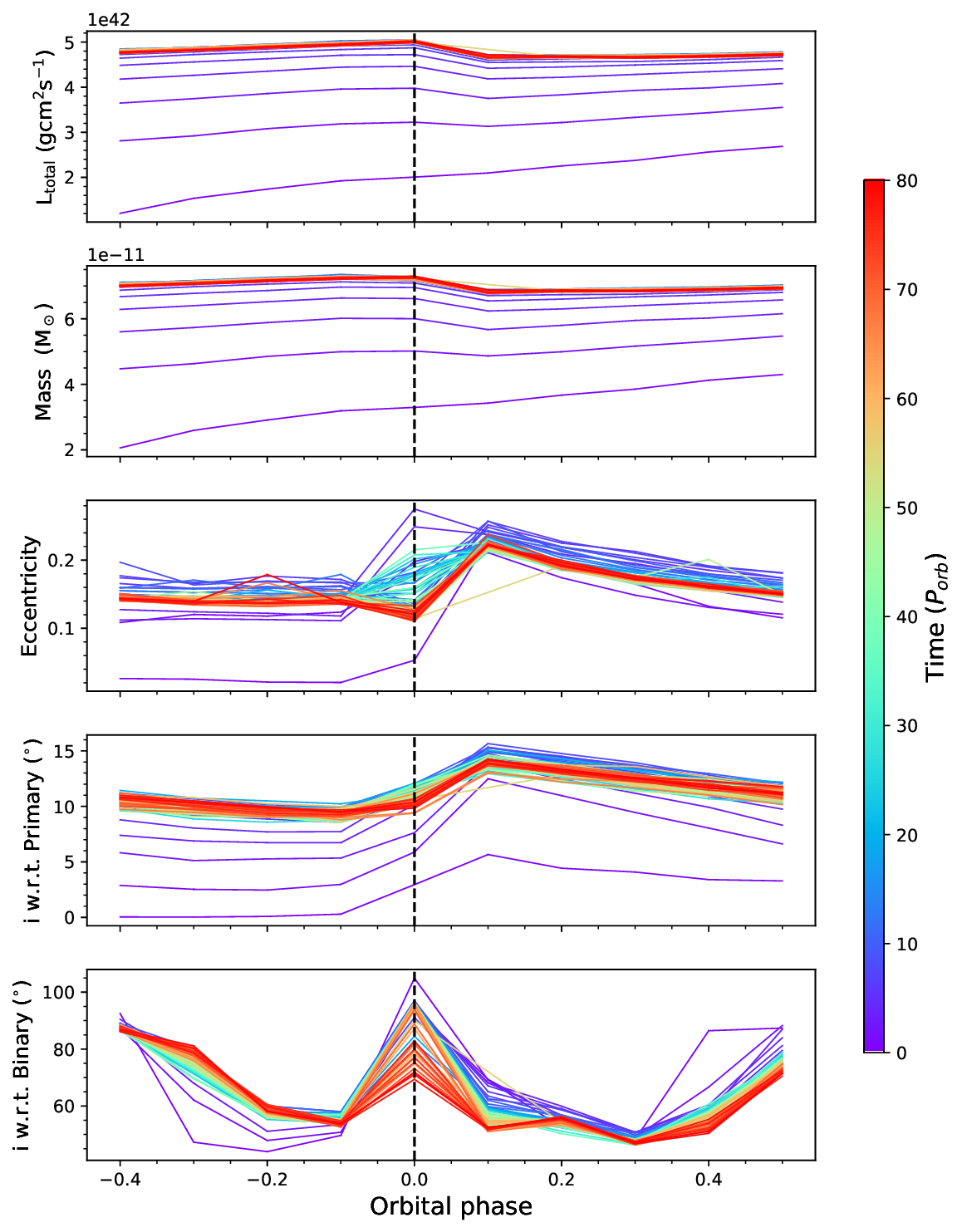}
    \caption{Same as Fig.~\ref{fig:pro_coplanar_phased}, but for the 135$^{\circ}$ simulation.} 
    \label{fig:retro_45_phased}
\end{figure}

\begin{figure}
	\includegraphics[width=\columnwidth]{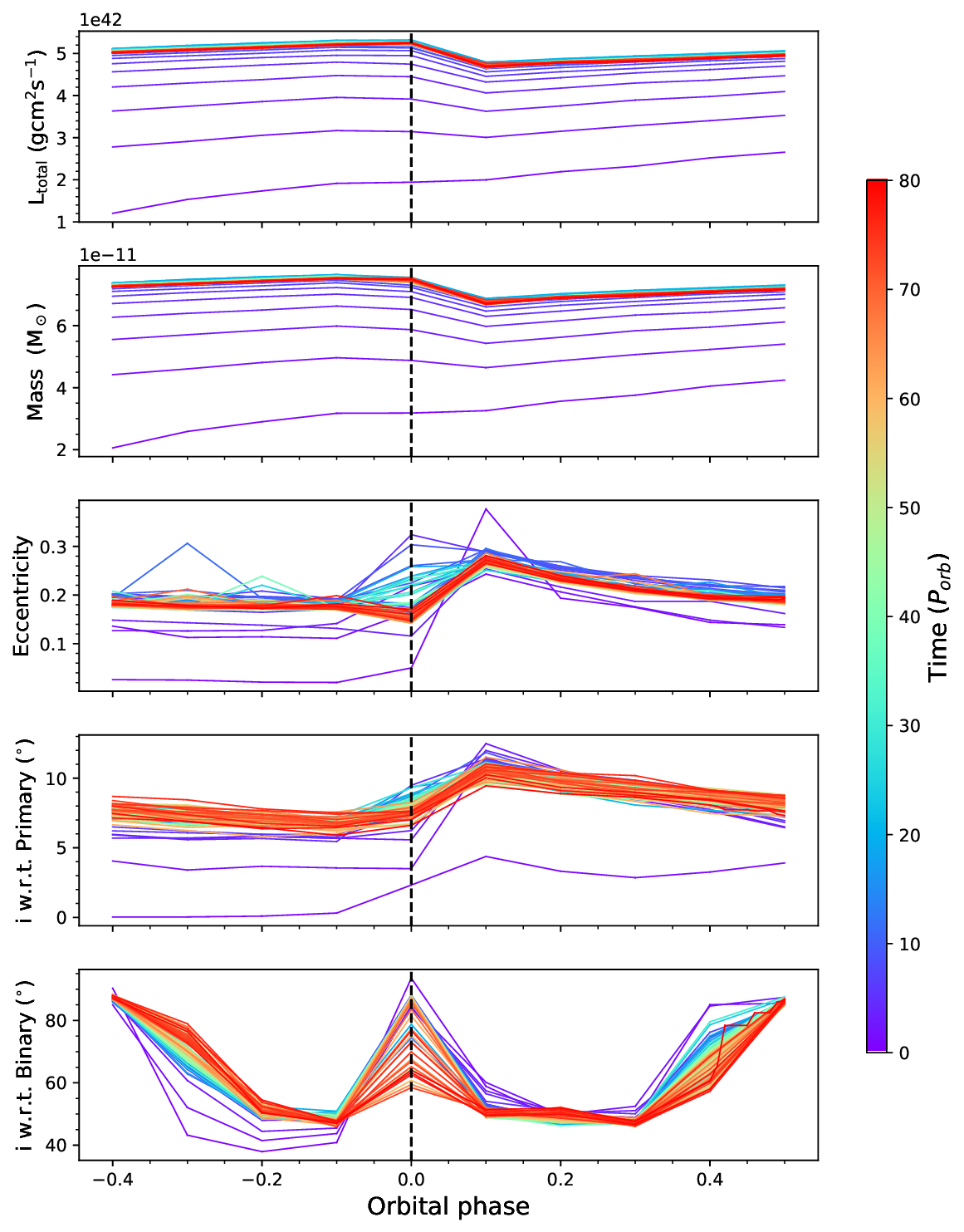}
    \caption{Same as Fig.~\ref{fig:pro_coplanar_phased}, but for the 120$^{\circ}$ simulation.} 
    \label{fig:retro_60_phased}
\end{figure}


\bsp	
\label{lastpage}
\end{document}